\documentclass{tlp}
\newtheorem{theorem}{Theorem}[section]
\newtheorem{cor}[theorem]{Corollary}
\newtheorem{lem}[theorem]{Lemma}
\newtheorem{prop}[theorem]{Proposition}
\newtheorem{definition}{Definition}[section]
\newtheorem{example}{Example}[section]
\newcommand{\pf}[1]{\langle\,#1\,\rangle}
\begin{document}
\def\baselinestretch{1.2}
\title[An abductive framework for computing knowledge base updates]
{An abductive framework for computing\\ knowledge base updates
\footnote{This paper is a revised and extended version of \cite{SI99}.}}
\author[C. Sakama and K. Inoue] 
{CHIAKI SAKAMA\\
Department of Computer and Communication Sciences\\ 
Wakayama University, Wakayama 640 8510, Japan\\
\email{sakama@sys.wakayama-u.ac.jp} 
\and KATSUMI INOUE\\
Department of Electrical and Electronics Engineering\\
Kobe University, Kobe 657 8501, Japan\\
\email{inoue@eedept.kobe-u.ac.jp}}
\date{}
\maketitle

\begin{abstract}\noindent 
This paper introduces an abductive framework for updating 
knowledge bases represented by extended disjunctive programs. 
We first provide a simple transformation from abductive programs to 
{\em update programs\/} which are logic programs specifying 
changes on abductive hypotheses.  
Then, {\em extended abduction}, which was introduced by the same authors 
as a generalization of traditional abduction, 
is computed by 
the answer sets of update programs. 
Next, different types of updates, 
{\em view updates\/} and {\em theory updates\/} 
are characterized by abductive programs and computed by update programs. 
The task of {\em consistency restoration\/} is also realized 
as special cases of these updates. 
Each update problem is comparatively assessed from the computational 
complexity viewpoint. 
The result of this paper provides a uniform framework for different types 
of knowledge base updates, and each update is computed using existing 
procedures of logic programming. \\


Submitted: Aug 20, 2001, revised: Jan 16, 2002 and June 25, 2002.

\end{abstract}

\section{Introduction} \label{sec:intro}

\subsection{Knowledge base updates} \label{sec:1.1}

When new information arrives at a knowledge base, 
an intelligent agent adjusts its current knowledge or belief 
to conform to the new circumstances. 
The problem of knowledge base updates is then how to specify the desired 
change in a knowledge base and to compute it automatically. 
The issue has been extensively studied in the context of 
databases and artificial intelligence (AI) 
and several different types of updates are studied in the literature. 
Among others, the following three cases are typical problem settings 
in database and knowledge base updating. 
The first case considers a knowledge base which contains two different kinds 
of knowledge, variable knowledge and invariable knowledge.  
In this case, updates are permitted only on the variable knowledge.  
Updates on the invariable part are then translated into 
updates on the variable part. 
An example of this type of updates is a {\em view update\/} in 
{\em deductive databases}, e.g., \cite{Dec90,KM90,GuL90,TO95}. 
A deductive database consists of invariable derivation rules 
(called an {\em intensional database\/}) and variable base facts 
(called an {\em extensional database\/}). 
Then, the view update problem in deductive databases is concerned with the 
problem of translating an update request on the derived facts into 
updates on the base facts. 
(For updating deductive databases, an excellent survey is in \cite{Dec98}.)  

In the second case, on the other hand, 
there is no distinction between variable and 
invariable knowledge, and the whole knowledge base is subject to change. 
In this case, an update is done by directly introducing 
new information to a knowledge base.  
When there are conflicts between the current knowledge and the new knowledge, 
a higher priority is put on the new one to produce a consistent theory 
as a whole.  This type of updates frequently appears in AI 
in the context of {\em theory updates\/} or {\em belief updates}, e.g., 
\cite{FUV83,Win90,KM91}.  
On the other hand, a knowledge base happens to be inconsistent in the face of 
contradictory knowledge. 
The third case handles {\em consistency restoration\/} in such 
knowledge bases. 
There are different sources which may cause inconsistency, 
e.g., conflicting information, violation of integrity constraints, etc. 
In this case, a knowledge base must be updated to restore consistency 
by detecting the source of inconsistency and repairing it. 
The problems of {\em integrity maintenance\/} in databases, 
e.g., \cite{TO95,Dec96}, 
and {\em inconsistency removal\/} in knowledge bases, 
e.g., \cite{PAA91,Ino94}, are of this kind. 

These three types of updates are not necessarily independent and orthogonal. 
In fact, integrity maintenance is often done as a subtask of a view update to 
remove contradiction derived by integrity constraints, and 
inconsistency removal is characterized as a special case of 
theory update which changes an inconsistent program to a consistent one. 
On the other hand, view updates and theory updates have been relatively 
independently studied so far and little connection exists between them. 
When a knowledge base is represented by a logic program, 
view updates are the problem of updating derived facts from a program, 
while theory updates are the problem of updating rules/facts included 
in a program.  
Thus, view updates and theory updates have seemingly different 
problem settings and goals. 
In fact, there are many studies which deal with updates in logic programming 
and deductive databases, while many of them are individual techniques to 
realize either view updates or theory updates. 
As far as the authors know, no study formalizes these two update problems 
in a single uniform framework. 

\subsection{Extended abduction} \label{sec:1.2}

{\em Abduction\/} is a form of hypothetical reasoning in AI. 
A traditional logical framework of abduction \cite{Poo88,KKT98} defines 
an {\em explanation\/} of a given observation as a set of hypotheses which, 
together with the background theory, logically entails the observation.  
More precisely, given a first-order theory $K$ and an observation $G$, 
traditional abduction computes a set $E$ of hypotheses satisfying 
\[K\cup E\models G\]
where $K\cup E$ is consistent.  

When a background knowledge base $K$ is {\em nonmonotonic}, however, 
the above framework of abduction is not sufficiently expressive.  
For example, consider the knowledge base written in a normal logic program:
\begin{eqnarray*}
K: && flies(x)\leftarrow bird(x), not\>ab(x),\\
   && ab(x)\leftarrow broken\mbox{-}wing(x),\\
   && bird(tweety)\leftarrow\mbox{},\\
   && bird(opus)\leftarrow\mbox{},\\
   && broken\mbox{-}wing(tweety)\leftarrow\mbox{}, 
\end{eqnarray*}
where $not$ represents {\em negation as failure}. 
If we observe that $tweety$ flies, there is a good reason 
to assume that the wound has already healed. 
Then, removing the fact $broken\mbox{-}wing(tweety)$ from 
the program explains the observation $flies(tweety)$.  
On the other hand, suppose that 
we later notice that $opus$ does not fly anymore. 
Since $flies(opus)$ is entailed by $K$, we now have to revise 
the knowledge base to block the derivation of $flies(opus)$ 
by assuming, for instance, $broken\mbox{-}wing(opus)$.  

Traditional abduction has difficulty to cope with these situations. 
First, abduction computes facts which are to be introduced to a program 
to explain an observation. However, abduction 
cannot compute facts which are to be removed from a 
program to explain an observation. 
Second, abduction computes explanations accounting for an observation, 
while it cannot compute hypotheses to {\em un\/}explain a phenomenon 
that does not hold anymore. 
To cope with the first problem, Inoue and Sakama \shortcite{IS95} introduce 
the notion of ``negative explanations''. 
Given a background knowledge base $K$ and an observation $G$, 
a set $F$ of hypotheses is called a {\em negative explanation\/} of $G$ if 
\[ K\setminus F\models G \]
where $K\setminus F$ is consistent. 
An explanation $E$ satisfying $K\cup E\models G$ is then 
called a {\em positive explanation}. 
On the other hand, the notion of ``anti-explanations'' is introduced 
to characterize the second situation. 
Given a background knowledge base $K$ and an observation $G$, a set $E$ of 
hypotheses is called a {\em (positive) anti-explanation\/} of $G$ if 
\[\, K\cup E\not\models G, \,\]
and a set $F$ of hypotheses is called a {\em negative anti-explanation\/} 
of $G$ if 
\[\, K\setminus F\not\models G. \,\]

These extensions of traditional abduction are called 
{\em extended abduction\/} \cite{IS95}. 
Extended abduction is particularly useful when a knowledge base is 
{\em nonmonotonic}.  In nonmonotonic theories, 
deletion of formulas may introduce new formulas. 
Thus, positive and negative explanations play a complementary role 
in accounting for an observation in nonmonotonic theories. 
On the other hand, anti-explanations are useful to account for 
{\em negative observations\/} which do not hold. 
In this respect, traditional abduction is concerned with explaining 
{\em positive observations\/} only.  
Negative observations are often perceived in real-life situations, and are 
analogous to the concept of {\em negative examples\/} in 
{\em inductive concept-learning}. 
Thus, anti-explanations play a dual role to explanations. 
Moreover, extended abduction not only enhances reasoning ability of 
traditional abduction, but has useful applications for 
nonmonotonic theory change \cite{IS95}, 
system repair problems \cite{BEGL97}, and incremental evolution of 
(inconsistent) requirement specifications \cite{NR99}. 

\subsection{The purpose of this paper} \label{sec:1.3}

The purposes of this paper are twofold. 
Our first goal is to provide a method of computing extended abduction. 
Many procedures exist for (traditional) abduction, while few is known 
for extended abduction with the exception of \cite{IS99}. 
Inoue and Sakama \shortcite{IS99} provide a computational method 
for extended abduction in a restricted class of normal logic program. 
By contrast, this paper considers extended abduction in 
{\em extended disjunctive programs\/} (EDPs) which are 
a fairly general class of logic programming. 
To compute extended abduction, this paper introduces an {\em update program\/} 
which is a logic program obtained by a simple program transformation. 
An update program specifies changes on abductive hypotheses, and 
(minimal) (anti-)explanations are computed by 
the {\em (U-minimal) answer sets\/} of an update program. 

Our second goal is to characterize various types of knowledge base updates 
through extended abduction. 
It is well known that knowledge base updates are related to abduction problems, 
and there are several studies which realize updates through abduction. 
However, due to the nature of traditional abduction, 
existing studies often adopt somewhat indirect formulations for representing 
hypotheses removal or view deletion 
(see Section~\ref{sec:7.2} for detailed discussion). 
In this paper we use extended abduction and formalize different types of update 
problems such as view updates, theory updates, and consistency restoration. 
These updates are then computed using update programs. 
We assess computational complexities and 
compare the difficulty of each update problem. 

This paper is a revised and extended version of \cite{SI99}. 
In the previous paper we considered knowledge base updates in 
{\em extended logic programs}.  In the present paper, 
we extend the techniques to extended disjunctive programs (EDPs) 
which possibly contain disjunction in a program. 
EDPs are strictly more expressive than extended (or normal) logic programs 
without disjunction, and are useful to express many practical problems in 
the complexity class $\Sigma^P_2$ \cite{EGM97}. 
In the context of updating data/knowledge bases, 
there are few studies which handle updating disjunctive (deductive) databases. 
The present paper is thus intended to provide a framework for 
(extended) abduction and update, which is applicable to 
a broader class of logic programming and deductive databases. 

The rest of this paper is organized as follows.  
Section~2 introduces a theoretical framework used in this paper. 
Section~3 introduces the notion of update programs 
and a method of computing extended abduction. 
Section~4 and Section~5 respectively characterize view updates and theory 
updates through extended abduction, and provide their computational methods 
by update programs.  Consistency restoration is also characterized as 
a special case of each update. 
Section~6 analyzes computational complexities of various update problems. 
Section~7 presents detailed comparisons with related work, 
and Section~8 concludes the paper.  

\section{Preliminaries} \label{sec:2}

\subsection{Extended disjunctive programs} \label{sec:2.1}

In this paper we consider knowledge bases represented as  
{\em extended disjunctive programs\/} (EDPs). 

An EDP is a set of {\em rules\/} of the form: 
\[ \mbox{ }\;\; L_1;\cdots; L_l\leftarrow\, L_{l+1},\,\ldots,\,L_m,\,not\,L_{m+1},\,
\ldots,\,not\,L_n\;\;\; (n\geq m\geq l\geq 0) \;\;\;\;\;(\dag)\]
where each $L_i$ is a literal, ``$;$'' represents ``or'', and 
$not$ represents {\em negation as failure} (NAF). 
$not\,L$ is also called an {\em NAF-literal}. 
The part left of $\leftarrow$ is the {\em head\/} and 
the part right of $\leftarrow$ is the {\em body\/} of the rule.  
We often use the Greek letter $\Sigma$ (resp.\ $\Gamma$) to represent 
the disjunction (resp.\ conjunction) in the head (resp.\ body). 
$\Sigma$ or $\Gamma$ is identified with the set of (NAF-)literals 
included in it. 
A rule is {\em disjunctive\/} if its head contains more than one literal.  
The head is possibly empty and 
a rule with the empty head is called an {\em integrity constraint}. 
A disjunctive rule with the empty body is called a {\em disjunctive fact}. 
A disjunctive fact $L_1;\cdots; L_l\leftarrow$ 
is simply written as $L_1;\cdots; L_l$. 
In particular, the non-disjunctive fact $L\leftarrow\mbox{}$ is 
identified with the literal $L$ and is simply called a {\em fact}. 
An EDP is called an {\em extended logic program\/} (ELP) 
if $l\le 1$ for each rule~($\dag$); and 
a {\em normal disjunctive program\/} (NDP) if every $L_i$ is an atom. 
An NDP is called a {\em normal logic program\/} (NLP) 
if $l\le 1$ for each rule~($\dag$). 
In this paper, a {\em program\/} means an EDP unless stated otherwise. 
A program (rule, (NAF-)literal) is {\em ground\/} if it contains no variable. 
A program $P$ is semantically identified with its ground instantiation, i.e., 
the set of all ground rules obtained from $P$ by substituting variables in 
$P$ by elements of its Herbrand universe in every possible way. 
Thus, a program containing variables is considered as 
a shorthand of its ground instantiation. 

The semantics of EDPs is given by the {\em answer set semantics\/} 
\cite{GL91}.  Let ${\cal L}_P$ be the set of all ground literals in the 
language of a program $P$. 
A set $S(\subseteq {\cal L}_P)$ {\em satisfies\/} the ground rule of 
the form~($\dag$) 
if $\{L_{l+1},\ldots,L_m\}\subseteq S$ and 
$\{\,L_{m+1},\ldots,L_n\,\}\cap S=\emptyset$ imply $L_i\in S$ for some 
$i\;(1\leq i\leq l)$.  
In particular, $S$ satisfies the ground integrity constraint 
$\leftarrow\, L_1,\,\ldots,\,L_m,\,not\,L_{m+1},\,\ldots,\,not\,L_n$ 
if $\{L_1,\ldots,L_m\}\not\subseteq S$ or 
$\{\,L_{m+1},\ldots,L_n\,\}\cap S\neq\emptyset$. 
Let $P$ be a $not$-free EDP (i.e., $m=n$ for each rule of~($\dag$)). 
Then, a set $S(\subseteq {\cal L}_P)$ 
is an {\em answer set\/} of $P$ if $S$ is a minimal set such that 
\begin{enumerate}
\item $S$ satisfies every ground rule from the ground instantiation of $P$, 
\item If $S$ contains a pair of complementary literals $L$ and $\neg L$, then $S={\cal L}_P$. 
\end{enumerate}
Next, let $P$ be any EDP and $S\subseteq {\cal L}_P$. 
Then, the $not$-free EDP $P^S$ is defined as follows: 
for every ground rule~($\dag$) from the ground instantiation of $P$, 
the rule $L_1;\cdots; L_l\leftarrow L_{l+1},\,\ldots,\,L_m$ is in $P^S$ 
if $\{L_{m+1},\ldots,L_n\}\cap S=\emptyset$. 
Then, $S$ is an {\em answer set\/} of $P$ if $S$ is an answer set of $P^S$. 
An EDP has none, one, or multiple answer sets in general. 
Answer sets coincide with {\em stable models\/} \cite{GL88} 
when $P$ is an NDP or an NLP. 

An answer set is {\em consistent\/} if it is not ${\cal L}_P$.
A program $P$ is {\em consistent\/} if it has a consistent answer set;
otherwise $P$ is {\em inconsistent}.
If a rule $R$ is satisfied in every answer set of $P$, 
it is written as $P\models R$. 
In particular, $P\models L$ if a literal $L$ is included in 
every answer set of $P$. 
When $P$ is inconsistent, we write $P\models \bot$ where $\bot$ is 
the reserved proposition in ${\cal L}_P$. 

\subsection{Abductive programs} \label{sec:2.2}

The abductive framework considered in this paper is based on 
{\em extended abduction\/} introduced by Inoue and Sakama \shortcite{IS95}. 

An {\em abductive program\/} is a pair $\pf{P,{\cal A}}$ where $P$ 
and ${\cal A}$ are EDPs. 
Every element in ${\cal A}$ is called an {\em abducible}. 
An abducible $A\in {\cal A}$ is also called an {\em abducible rule\/} 
(resp.\ {\em abducible fact\/}) if $A$ is a rule (resp. a fact). 
An abducible containing variables is considered as 
a shorthand of its ground instantiation. 
So any instance $A$ of an element from ${\cal A}$ is also an 
abducible and is written as $A\in {\cal A}$. 
Abducibles are hypothetical rules which are used to account for 
an observation together with the background knowledge $P$. 
Similar frameworks are also introduced in \cite{Ino94,IS98}. 
An abductive program $\pf{P,{\cal A}}$ is {\em consistent\/} if $P$ is 
consistent. 
Without loss of generality, we assume that for any rule 
$\Sigma\leftarrow\Gamma$ from $P$, 
$\Sigma\cap {\cal A}\neq\emptyset$ implies both 
$\Sigma\subseteq {\cal A}$ and $\Gamma=\emptyset$.%
\footnote{We pose this assumption just by technical reasons. 
A similar assumption is assumed, for instance, in \cite{KKT98}.} 
If there is a rule $\Sigma\leftarrow\Gamma$ with 
$\Sigma\cap {\cal A}\neq\emptyset$ such that 
$\Sigma\not\subseteq {\cal A}$ or $\Gamma\neq\emptyset$, 
then any $A\in\Sigma\cap {\cal A}$ is made a non-abducible by 
introducing a rule $A\leftarrow A'$ with a new abducible $A'$ and 
replacing $A$ with $A'$ in every (disjunctive) fact consisting 
abducibles only. 

We also assume that for any disjunctive fact $\Sigma\leftarrow$ from $P$, 
$\Sigma\subseteq {\cal A}$ implies $\Sigma\in {\cal A}$. 
That is, if a program contains a disjunctive fact $\Sigma$ which consists of 
abducibles, $\Sigma$ itself is included in ${\cal A}$ as an abducible. 
This condition is natural, since any disjunctive fact in $P$ 
which consists of abducibles is considered a hypothesis.
On the other hand, any disjunctive fact which is not included in $P$ 
is freely specified in ${\cal A}$ as a possible hypothesis. 

Let $\pf{P,{\cal A}}$ be an abductive program and $G$ a ground literal 
representing a {\em positive observation}. 
A pair $(E,F)$ is a {\em skeptical explanation\/} 
of $G$ with respect to $\pf{P,{\cal A}}$ if 
\begin{enumerate}
\item $(P\setminus F)\cup E\models G$, 
\item $(P\setminus F)\cup E$ is consistent, 
\item $E\subseteq {\cal A}\setminus P$ and $F\subseteq {\cal A}\cap P$. 
\end{enumerate}
If the first condition is replaced by 
``$G$ is true in {\em some\/} answer set of $(P\setminus F)\cup E$'', 
$(E,F)$ is called a {\em credulous explanation}.  
Any skeptical explanation is a credulous explanation. 
On the other hand, given a ground literal $G$ representing a 
{\em negative observation}, 
a pair $(E,F)$ is a {\em credulous anti-explanation\/} 
of $G$ with respect to $\pf{P,{\cal A}}$ if 
\begin{enumerate}
\item $(P\setminus F)\cup E\not\models G$, 
\item $(P\setminus F)\cup E$ is consistent, 
\item $E\subseteq {\cal A}\setminus P$ and $F\subseteq {\cal A}\cap P$. 
\end{enumerate}
If the first condition is replaced by 
``$G$ is true in no answer set of $(P\setminus F)\cup E$'', 
$(E,F)$ is called a {\em skeptical anti-explanation}.  
Any skeptical anti-explanation is a credulous anti-explanation. 
In particular, when $G=\bot$, 
the first and the second conditions are identical. 
In this case, the credulous anti-explanation $(E,F)$ of $\bot$ is a 
hypothesis which turns a (possibly inconsistent) $P$ 
to a consistent program $(P\setminus F)\cup E$. 

Throughout the paper, a skeptical/credulous (anti-)explanation is simply 
called an (anti-)explanation when such a distinction is not important. 
A positive or negative observation is also simply called an observation 
when no confusion arises.  
Without loss of generality, an observation is assumed to be 
a (non-abducible) ground literal \cite{IS96}. 
By the third condition, 
the introduced hypotheses $E$ are abducibles which are not included in 
the program $P$, while the removed hypotheses $F$ are abducibles which are 
included in $P$. Thus, it holds that $E\cap F=\emptyset$ for any 
(anti-)explanation $(E,F)$. 
Among (anti-)explanations, {\em minimal (anti-)explanations\/} are of 
particular interest. 
An (anti-)explanation $(E,F)$ of an observation $G$ 
is called {\em minimal\/} if for any 
(anti-)explanation $(E',F')$ of $G$, 
$E'\subseteq E$ and $F'\subseteq F$ imply $E'=E$ and $F'=F$.  

Note that the abduction problem considered here is different from the usual 
one based on traditional normal abduction \cite{KKT98}.%
\footnote{To distinguish extended abduction from traditional one, 
we call traditional abduction {\em normal abduction}, hereafter.} 
That is, given an abductive program $\pf{P,{\cal A}}$, 
{\em normal abduction\/} computes a {\em skeptical explanation\/} 
(resp.\ {\em credulous explanation\/}) 
$E$ of a positive observation $G$ satisfying 
\begin{enumerate}
\item $P\cup E\models G$\hspace{3mm} (resp.\ $G$ is true in some answer set of $P\cup E$), 
\item $P\cup E$ is consistent, 
\item $E\subseteq {\cal A}\setminus P$. 
\end{enumerate}
Compared with normal abduction, extended abduction abduces hypotheses 
which are not only introduced to a program 
but also removed from a program to explain observations. 
Moreover, anti-explanations are used to {\em un\/}explain a 
negative observation which is not true. 
With this respect, normal abduction is considered as a special case of 
extended abduction where only hypotheses introduction is considered for 
explaining positive observations. 
 
In an abductive program $\pf{P,{\cal A}}$, 
$P$ and ${\cal A}$ are semantically identified with their ground 
instantiations, so that set operations over them are defined on 
the ground instances. 
Thus, when $(E,F)$ contains variables, 
$(P\setminus F)\cup E$ 
means that deleting every instance of $F$ from $P$ and 
adding any instance of $E$ to $P$. 
Also, when $E$ contains variables, 
the set inclusion $E'\subseteq E$ is defined for any instance $E'$ of $E$.  
Generally, given sets $S$ and $T$ of literals/rules containing variables, 
any set operation $\circ$ is defined as $S\circ T=inst(S)\circ inst(T)$ 
where $inst(S)$ is the ground instantiation of $S$  \cite{Ino00}. 
For example, when $p(x)\in T$, 
for any constant ``$a$'' in the language of $T$, 
it holds that $\{p(a)\}\subseteq T$, 
$\{p(a)\}\setminus T=\emptyset$, and 
$T\setminus\{p(a)\}=(T\setminus\{p(x)\})\,\cup\,\{\,p(y)\,\mid\,y\neq a\}$, 
and so on. 
Also, any literal/rule in a set is identified with its 
variant modulo variable renaming. 

\begin{example} \label{var-ex}
Let $\pf{P,{\cal A}}$ be the abductive program such that 
\begin{eqnarray*}
P:&& g\leftarrow p(x),\, not\,r\\
&& r\leftarrow q(a),\\
&& q(a)\leftarrow,\;\; q(b)\leftarrow.\\
{\cal A}:&& p(x),\; q(x). 
\end{eqnarray*}
Then, $(\{p(x)\},\{q(x)\})$ is a skeptical explanation of $g$, while 
$(\{p(a)\},\{q(a)\})$ and $(\{p(b)\},\{q(a)\})$ are the 
minimal skeptical explanations of $g$.  
\end{example}

Suppose an abductive program $\pf{P,{\cal A}}$ where 
${\cal A}$ contains rules or disjunctive facts.  In this case, 
$\pf{P,{\cal A}}$ is transformed to a semantically equivalent abductive 
program in which abducibles contain only (non-disjunctive) facts as follows. 
Given an abductive program $\pf{P,{\cal A}}$, let 
\[{\cal R}=\{\,\Sigma\leftarrow\Gamma\,\mid\, (\Sigma\leftarrow\Gamma)\in {\cal A}\:\:\mbox{and}\:\: \Sigma\leftarrow\Gamma\:\mbox{is not a non-disjunctive fact}\,\}\,.\]
Then, we define  
\begin{eqnarray*}
P^{\rm n}&=&(P\setminus {\cal R})\;\cup\;
\{\,\Sigma\leftarrow\Gamma,\gamma_R\,\mid\,R=(\Sigma\leftarrow\Gamma)\in {\cal R}\,\}\\
 && \mbox{}\;\;\;\;\;\;\;\;\;\;\;\;\;\;\cup\;\{\, \gamma_R\leftarrow\,\mid\, R\in {\cal R}\cap P\,\},\\
{\cal A}^{\rm n}\,&=&\,({\cal A}\setminus {\cal R})\;\cup\;\{\,\gamma_R\,\mid\, R\in {\cal R}\,\}, 
\end{eqnarray*}
where $\gamma_R$ is a newly introduced atom (called the {\em name\/} of $R$) 
uniquely associated with each rule $R$ in ${\cal R}$. 
For any rule $R\in {\cal R}$, 
we refer to its name using the function $n(R)=\gamma_R$. 
In particular, we define that any abducible fact $L\leftarrow$ has the 
name $L$, i.e., $n(L)=L$. 
We call $\pf{P^{\rm n},{\cal A}^{\rm n}}$ the {\em normal form\/} of 
$\pf{P,{\cal A}}$.  With this setting, for any observation $G$ 
there is a 1-1 correspondence between 
(anti-)explanations with respect to $\pf{P,{\cal A}}$ 
and those with respect to $\pf{P^{\rm n},{\cal A}^{\rm n}}$. 
In what follows, $n(E)=\{\,n(R)\,\mid\,R\in E\,\}$. 

\begin{prop}[normal form transformation]  \label{naming-prop}
Let $\pf{P,{\cal A}}$ be an abductive program 
and $\pf{P^{\rm n},{\cal A}^{\rm n}}$ its normal form. 
Then, an observation $G$ has a (minimal) credulous/skeptical 
(anti-)explanation $(E,F)$ with respect to $\pf{P,{\cal A}}$ 
iff $G$ has a (minimal) credulous/skeptical 
(anti-)explanation $(n(E),n(F))$ 
with respect to $\pf{P^{\rm n},{\cal A}^{\rm n}}$. 
\begin{proof}
By the definition of $\pf{P^{\rm n},{\cal A}^{\rm n}}$, 
$G$ is included in a consistent answer set of $(P\setminus F)\,\cup\,E$ iff 
$G$ is included in a consistent answer set of 
$(P^{\rm n}\setminus n(F))\,\cup\,n(E)$ 
with $n(E)\subseteq {\cal A}^{\rm n}\setminus P^{\rm n}$ 
and $n(F)\subseteq {\cal A}^{\rm n}\cap P^{\rm n}$. 
Hence, the result holds. 
\end{proof}
\end{prop}

\begin{example}\rm \label{ex-abrule}
Let $\pf{P,{\cal A}}$ be the abductive program such that 
\begin{eqnarray*}
P: && flies(x)\leftarrow bird(x),\\
   && bird(x)\leftarrow penguin(x),\\
   && bird(polly)\leftarrow,\\
   && penguin(tweety)\leftarrow.\\
{\cal A}: && flies(x)\leftarrow bird(x),\\
&& \neg flies(x)\leftarrow penguin(x). 
\end{eqnarray*}
Then, the positive observation $G=\neg flies(tweety)$ has the minimal 
skeptical explanation 
$(E,F)=(\{\,\neg flies(tweety)\leftarrow penguin(tweety)\,\},
\{\,flies(tweety)\leftarrow bird(tweety)\,\})$. 

On the other hand, the abductive program $\pf{P,{\cal A}}$ is transformed 
to the normal form $\pf{P^{\rm n},{\cal A}^{\rm n}}$ where 
\begin{eqnarray*}
P^{\rm n}: && flies(x)\leftarrow bird(x),\,\gamma_1(x),\\
   && bird(x)\leftarrow penguin(x),\\
   && \neg flies(x)\leftarrow penguin(x),\,\gamma_2(x),\\
   && \gamma_1(x)\leftarrow,\;\;\; bird(polly)\leftarrow,\\
   && penguin(tweety)\leftarrow,\\
{\cal A}^{\rm n}:&& \gamma_1(x),\, \gamma_2(x). 
\end{eqnarray*}
Here, $\gamma_1(x)$ and $\gamma_2(x)$ are the names of 
the rules $flies(x)\leftarrow bird(x)$ and 
$\neg flies(x)\leftarrow penguin(x)$, respectively. 
In this program, $G=\neg flies(tweety)$ has the minimal skeptical 
explanation $(\{\,\gamma_2(tweety)\,\}$, $\{\,\gamma_1(tweety)\,\})$, 
which corresponds to the minimal explanation $(E,F)$ presented above. 

Note that $(E',F')=(\{\,\neg flies(x)\leftarrow penguin(x)\,\},
\{\,flies(x)\leftarrow bird(x)\,\})$ is also an explanation of $G$ 
with respect to $\pf{P,{\cal A}}$, 
but it is not minimal (cf. Example~\ref{var-ex}). 
In fact, $E\subseteq E'$ and $F\subseteq F'$. 
\end{example} 

By the definition of abductive programs, 
a program includes no disjunctive rule which contains both abducibles 
and non-abducibles in its head. 
Thus, if there is a disjunctive fact $\Sigma\leftarrow$ in $P$, 
every disjunct in $\Sigma$ is an abducible.  This justifies the 
replacement of the disjunction $\Sigma$ with a new abducible $\gamma$ 
in the normal form. 

\begin{example} 
Let $\pf{P,{\cal A}}$ be the abductive program such that 
\begin{eqnarray*}
P: &&  p\leftarrow a\,,\;\;\; \\
&& p\leftarrow b\,,\;\;\; \\
&& a\,; b\leftarrow\,.\\
{\cal A}: && a,\; b,\; (a\,; b).  
\end{eqnarray*}
Transform $\pf{P,{\cal A}}$ to $\pf{P^{\rm n},{\cal A}^{\rm n}}$ with 
\begin{eqnarray*}
P^{\rm n}: &&  p\leftarrow a\,,\\
&& p\leftarrow b\,,\\
&& a\,; b\leftarrow \gamma\,,\\
&& \gamma\leftarrow.\\
{\cal A}^{\rm n}: && a,\; b,\; \gamma.
\end{eqnarray*}
Then, the negative observation $p$ has the skeptical anti-explanation 
$(\emptyset,\{\gamma\})$ with respect to $\pf{P^{\rm n},{\cal A}^{\rm n}}$, 
which corresponds to the anti-explanation 
$(\emptyset,\{a;b\})$ with respect to $\pf{P,{\cal A}}$. 
\end{example}

Using the transformation, any abductive program having abducible rules is 
reduced to an abductive program having only (non-disjunctive) 
abducible facts. 
Thus, in the next section we consider an abductive program $\pf{P,{\cal A}}$ 
where ${\cal A}$ contains only (non-disjunctive) facts, 
unless specified otherwise.%
\footnote{By contrast, \cite{IS02} introduces a method of directly computing 
(anti-)explanations which are disjunctions of abducibles.} 

\section{Extended abduction through update programs} \label{sec:3}

In this section we introduce the notion of update programs 
and characterize extended abduction through them. 

\subsection{Update programs} \label{sec:3.1} 

Suppose an abductive program $\pf{P,{\cal A}}$ where 
${\cal A}$ consists of abducible facts. 
Then, update rules/programs are defined as follows. 

\begin{definition}[update rules]\label{def-ur}
Given an abductive program $\pf{P,{\cal A}}$, 
the set $UR$ of {\em update rules\/} is defined as follows. 
\begin{enumerate}
\item For any literal $a\in {\cal A}$, 
the following rules are in $UR$: 
\begin{eqnarray*}
&& a\leftarrow not\,\overline{a},\\
&& \overline{a}\leftarrow not\,a,
\end{eqnarray*}
where $\overline{a}$ is a newly introduced atom uniquely associated with $a$. 
For notational convenience, the above pair of rules is 
expressed as $abd(a)$, hereafter. 
\item For any literal $a\in {\cal A}\setminus P$, 
the following rule is in $UR$: 
\[ +a\leftarrow a\,. \]
\item For any literal $a\in {\cal A}\cap P$, 
the following rule is in $UR$: 
\[ -a\leftarrow not\,a\,. \]
\end{enumerate}
Here, $+a$ and $-a$ are atoms which are 
uniquely associated with any $a\in {\cal A}$. 
These are called {\em update atoms}. 
\end{definition}

By the definition, the atom $\overline{a}$ becomes true iff $a$ is not true. 
The pair of rules in $abd(a)$ then specify the situation that 
an abducible $a$ is true or not. 
Similar transformations are introduced in \cite{SI91,Ino94} 
in the context of transforming abductive programs to normal logic programs. 
The pair of rules in $abd(a)$ is also represented by 
the semantically equivalent disjunctive fact 
\[ a;\overline{a}\leftarrow. \]
This replacement is useful to avoid introducing unstratified negation in 
$abd(a)$ when the original program $P$ is stratified. 

In the second condition, 
when $p(x)\in {\cal A}$, $p(a)\in P$ and $p(t)\not\in P$ for 
$t\neq a$, the rule precisely becomes $+p(t)\leftarrow p(t)$ for any 
$t\neq a$.  In such a case, the rule is shortly written as 
$+p(x)\leftarrow p(x),\,x\neq a$.  Generally, the rule becomes 
$+p(x)\leftarrow p(x),\,x\neq t_1,\ldots,x\neq t_n$ for $n$ such instances. 
The rule $+a\leftarrow a$ derives the atom $+a$ 
if an abducible $a$ which is not in $P$ is to be true. 
In contrast, the rule $-a\leftarrow not\,a$ derives 
the atom $-a$ if an abducible $a$ which is in $P$ is not to be true. 
Thus, update atoms represent the change of truth values of 
abducibles in a program, i.e., $+a$ means the introduction of $a$, 
while $-a$ means the deletion of $a$. 
When an abducible $a$ contains variables, the associated update atom 
$+a$ or $-a$ is supposed to have exactly the same variables.  In this case, 
an update atom is semantically identified with its ground instances. 
The set of all update atoms associated with the abducibles in ${\cal A}$ 
is denoted by ${\cal UA}$. 
We define that ${\cal UA}={\cal UA}^+\cup {\cal UA}^-$, where 
${\cal UA}^+$ (resp.\ ${\cal UA}^-$) is 
the set of update atoms of the form $+a$ (resp.\ $-a$). 

\begin{definition}[update programs]\label{def-up}
Given an abductive program $\pf{P,{\cal A}}$, 
its {\em update program\/} $UP$ is defined as an EDP such that 
\[ UP=(P\setminus {\cal A})\,\cup\,UR\,.\]
\end{definition}
$UP$ becomes an ELP when $P$ is an ELP. 

\begin{definition}[U-minimal answer sets]
An answer set $S$ of $UP$ is called {\em U-minimal\/} 
if there is no answer set $T$ of $UP$ such that  
$T\cap {\cal UA}\subset S\cap {\cal UA}$. 
\end{definition}

By the definition, U-minimal answer sets exist whenever 
$UP$ has answer sets. 
A U-minimal answer set is used for characterizing a minimal change in $P$. 
In particular, when there is no observation, 
there is a 1-1 correspondence between the U-minimal answer sets of $UP$ and 
the consistent answer sets of $P$. 

\begin{prop}[U-minimal answer sets vs.\ answer sets]
Let $\pf{P,{\cal A}}$ be an abductive program and $UP$ its update program. 
Then, $P$ has a consistent answer set $T$ iff 
$UP$ has a U-minimal answer set $S$ such that 
$S\cap {\cal UA}=\emptyset$ and $S\cap {\cal L}_P=T$. 
\begin{proof}  
Let $T$ be a consistent answer set of $P$. 
Put $S=T\,\cup\,\{\,\overline{a}\,\mid a\in {\cal A}\setminus P\,\}$, 
then $S\cap {\cal L}_P=T$. 
By the definition of abductive programs, 
any abducible $a\in {\cal A}\setminus P$ does not appear in the head 
of any rule which is not a fact in $P$. 
So $T$ contains no abducible $a$ such that $a\in {\cal A}\setminus P$, 
then $\overline{a}\in S$ implies $a\not\in S$. 
Next, consider $UP^S=(P\setminus {\cal A})^S\,\cup\,UR^S$. 
It holds that $(P\setminus {\cal A})^S=(P\setminus {\cal A})^T=P^T\setminus {\cal A}$. 
For any $abd(a)\in UR$, 
$(a\leftarrow)\in UR^S$ iff $\overline{a}\not\in S$ iff 
$a\in {\cal A}\cap P$; and 
$(\overline{a}\leftarrow)\in UR^S$ iff $a\not\in S$ iff 
$a\in {\cal A}\setminus P$. 
Also, any $+a\leftarrow a$ in $UR$ is also in $UR^S$. 
Since $T$ is an answer set of $P$, by the construction of $S$ 
it contains every abducible $a$ such that $a\in {\cal A}\cap P$. 
Thus, any $-a\leftarrow not\,a$ in $UR$ is not included in $UR^S$. 
Hence, $UP^S=(P\setminus {\cal A})^S\,\cup\,UR^S
=(P^T\setminus {\cal A})\,
\cup\,\{\,a\leftarrow\,\mid\, a\in {\cal A}\cap P\,\}\,
\cup\,\{\,\overline{a}\leftarrow\,\mid\, a\in {\cal A}\setminus P\,\}\,
\cup\,\{\,+a\leftarrow~a\,\mid\, a\in {\cal A}\setminus P\,\}
=P^T\,\cup\,\{\,\overline{a}\leftarrow\,\mid\, \overline{a}\in S\,\}\,
\cup\,\{\,+a\leftarrow a\,\mid\, a\in {\cal A}\setminus P\,\}$. 
As $T$ is an answer set of $P^T$ and $a\not\in S$, 
$S$ becomes an answer set of $UP^S$. 
Thus, $S$ is an answer set of $UP$. 
Since $S\cap {\cal UA}=\emptyset$, $S$ is also U-minimal. 

Conversely, let $S$ be a U-minimal answer set of $UP$ such that 
$S\cap {\cal UA}=\emptyset$. 
By $S\cap {\cal UA}=\emptyset$, $S$ contains no literal in 
${\cal A}\setminus P$.  Hence, $S$ is a consistent answer set. 
Also, it implies 
$a\in S\cap {\cal A}$ iff $a\in {\cal A}\cap P$ iff $a\in {\cal A}\cap UP^S$. 
Put $T=S\cap {\cal L}_P$.  
Then, $P^T=\{\,\Sigma\leftarrow\Gamma\,\mid\,(\Sigma\leftarrow\Gamma)\in UP^S\,\mbox{ and } \Sigma\subseteq {\cal L}_P\,\}$.  
Since $S$ is a consistent answer set of $UP^S$, 
$T$ becomes a consistent answer set of $P^T$. 
Hence, $T$ is a consistent answer set of $P$. 
\end{proof}
\end{prop}

\begin{example}\label{trans-ex}
Let $\pf{P,{\cal A}}$ be the abductive program such that 
\begin{eqnarray*}
P: &&  p\leftarrow b\,,\\
&& q\leftarrow a,\, not\,b\,,\\
&& a\leftarrow\,.\\
{\cal A}: && a,\; b\,.
\end{eqnarray*}
Then, $UP$ becomes 
\begin{eqnarray*}
UP: &&  p\leftarrow b\,,\\ 
&& q\leftarrow a,\, not\,b\,,\\
&& abd(a),\;\;abd(b)\,,\\
&& -a \leftarrow not\,a\,,\\
&& +b \leftarrow b\,.
\end{eqnarray*}
Here, $UP$ has four answer sets: 
$S_1=\{\,a,b,+b,p\,\}$, $S_2=\{\,\overline{a},b,-a,+b,p\,\}$, 
$S_3=\{\,a,\overline{b},q\,\}$, and 
$S_4=\{\,\overline{a},\overline{b},-a\,\}$. 
Of these, $S_3$ is the U-minimal answer set 
and $S_3\cap {\cal L}_P$ coincides with the answer set of $P$. 
\end{example}

\subsection{Computing (anti-)explanations through UP}

Next, we provide a method of computing (anti-)explanations through 
update programs. 
A positive observation $G$ represents an evidence which is to be true 
in a program. The situation is specified by the integrity constraint 
\[ \leftarrow\, not\,G\,,\]
which represents that ``$G$ should be true''.  
By contrast, a negative observation $G$ represents an evidence which is 
not to be true in a program. The situation is specified 
by the integrity constraint 
\[ \leftarrow\, G\,,\] 
which represents that ``$G$ must not be true''.  

For instance, to explain the positive observation 
$p$ in the program $P$ of Example~\ref{trans-ex}, 
consider the program $UP\,\cup\,\{\,\leftarrow not\,p\,\}$. 
It has two answer sets: $S_1$ and $S_2$, 
of which $S_1$ is the U-minimal answer set. 
Observe that the positive observation 
$p$ has the unique minimal (skeptical) explanation $(\{b\},\emptyset)$ 
with respect to $\pf{P,{\cal A}}$. 
The situation is expressed by the update atom $+b$ in $S_1$. 
On the other hand, to unexplain the negative observation $q$ in $P$, 
consider the program $UP\,\cup\,\{\, \leftarrow q\,\}$. 
It has three answer sets: $S_1$, $S_2$, and $S_4$, of which 
$S_1$ and $S_4$ are the U-minimal answer sets. 
Here, the negative observation $q$ has two minimal (skeptical) 
anti-explanations $(\{b\},\emptyset)$ 
and $(\emptyset,\{a\})$ with respect to $\pf{P,{\cal A}}$. 
The situations are respectively 
expressed by the update atom $+b$ in $S_1$ and $-a$ in $S_4$. 
Note that when the positive observation $p$ and 
the negative observation $q$ are given 
at the same time, $S_1$ becomes the unique U-minimal answer set of 
$UP\,\cup\,\{\,\leftarrow not\,p\,\}\,\cup\,\{\,\leftarrow q\,\}$.%
\footnote{When there are positive observations 
$p_1,\ldots,p_m$ and negative observations $q_1,\ldots,q_n$, 
instead of considering the 
$(m+n)$-goals $\leftarrow not\,p_i$ and $\leftarrow q_j$, 
the same effect is achieved by introducing the rule 
$g\leftarrow p_1,\ldots,p_m,not\,q_1,\ldots,not\,q_n$ to $UP$ 
and considering the single goal $\leftarrow not\,g$.}

These examples illustrate that the U-minimal answer sets are used to 
compute minimal (anti-)explanations of extended abduction. 
Note that the constraint $\leftarrow not\,G$ extracts answer sets in 
which $G$ is true, but this does not imply that $G$ is true in every answer 
set of $(P\setminus F)\cup E$. 
To know that $(E,F)$ is a skeptical explanation of $G$, 
we need an additional test for checking the entailment of $G$ from 
$(P\setminus F)\cup E$. 

\begin{prop}[credulous vs.\ skeptical explanations]\label{pre-prop}
Let $\pf{P,{\cal A}}$ be an abductive program 
and $G$ a positive observation. 
Suppose that $(E,F)$ is a credulous explanation of $G$ with respect to 
$\pf{P,{\cal A}}$. 
Then, $(E,F)$ is a skeptical explanation of $G$ with respect to 
$\pf{P,{\cal A}}$ iff 
$(P\setminus F)\cup E\,\cup\,\{\,\leftarrow G\,\}$ is inconsistent.  
\begin{proof}
When $(E,F)$ is a credulous explanation of $G$ with respect to $\pf{P,{\cal A}}$, 
$(P\setminus F)\cup E$ has a consistent answer set in which $G$ is true. 
Then, $(E,F)$ is a skeptical explanation of $G$ with respect to $\pf{P,{\cal A}}$ \\
iff $(P\setminus F)\cup E$ has no consistent answer set in which $G$ is not true\\
iff $(P\setminus F)\cup E\,\cup\,\{\,\leftarrow G\,\}$ is inconsistent. 
\end{proof}
\end{prop}

\begin{example} \label{skpandcre-ex}
Let $\pf{P,{\cal A}}$ be the abductive program such that 
\begin{eqnarray*}
P: &&  p\,; q\leftarrow a\,,\;\;\; \\
&& \neg q\leftarrow not\,b\,,\;\;\; \\
&& b\leftarrow\,.\\
{\cal A}: && a,\; b\,.
\end{eqnarray*}
Given the positive observation $G=p$, 
$(E,F)=(\{a\},\{b\})$, $(\{a\},\emptyset)$ are two credulous explanations. 
Among them, 
$(\{a\},\{b\})$ is also the skeptical explanation of $G$ where 
$(P\setminus\{b\})\cup\{a\}\cup\{\,\leftarrow p\,\}$ is inconsistent. 
\end{example}

In what follows, given sets $E\subseteq {\cal A}$ and 
$F\subseteq {\cal A}$, we define 
$E^+=\{\,+a\mid a\in E\,\}$ and $F^-=\{\,-a\mid a\in F\,\}$. 
Conversely, given sets $E^+\subseteq {\cal UA}^+$ and 
$F^-\subseteq {\cal UA}^-$, we define 
$E=\{\,a\mid +a\in E^+\,\}$ and $F=\{\,a\mid -a\in F^-\,\}$. 
Then, (minimal) credulous/skeptical explanations are computed 
by update programs as follows. 

\begin{theorem}[computing credulous explanations through UP]\label{vu-lem}
Let $\pf{P,{\cal A}}$ be an abductive program, 
$UP$ its update program, and $G$ a positive observation. 
\begin{enumerate}
\item The pair $(E,F)$ is a credulous explanation of $G$ iff 
$UP\cup \{\,\leftarrow not\,G\,\}$ has a consistent answer set 
$S$ such that $E^+=S\cap {\cal UA}^+$ and $F^-=S\cap {\cal UA}^-$. 

\item The pair $(E,F)$ is a minimal credulous explanation of $G$ iff 
$UP\cup \{\,\leftarrow not\,G\,\}$ has a consistent U-minimal answer set 
$S$ such that $E^+=S\cap {\cal UA}^+$ and $F^-=S\cap {\cal UA}^-$. 
\end{enumerate}
\begin{proof}\rm 
1.  Let $S$ be a consistent answer set of $UP\cup \{\,\leftarrow not\,G\,\}$ 
such that $E^+=S\cap {\cal UA}^+$ and $F^-=S\cap {\cal UA}^-$. 
For each $+a\in E^+$ and $-b\in F^-$, $a\in S$ and $b\not\in S$ hold 
respectively.  Then, $a\leftarrow$ and $\overline{b}\leftarrow$ 
are respectively produced by $abd(a)$ and $abd(b)$ in $UP^S$, so that 
$(a\leftarrow)\in UP^S$ and $(b\leftarrow)\not\in UP^S$. 
By the definition, 
$+a\in E^+$ implies $a\in E$ and $-b\in F^-$ implies $b\in F$, 
so $UP^S$ contains a rule $\Sigma\leftarrow\Gamma$ with 
$\Sigma\subseteq {\cal L}_P$ 
iff $((P\setminus F)\cup E)^S$ has the same rule. 
Put $T=S\cap {\cal L}_P$.  
As $G\in S$, $T$ is a consistent answer set of 
$(P\setminus F)\cup E$ in which $G$ is true. 
Since $E\subseteq {\cal A}\setminus P$ and $F\subseteq {\cal A}\cap P$, 
$(E,F)$ is a credulous explanation of $G$. 
Conversely, suppose that $(E,F)$ is a credulous explanation of $G$. 
Then, there is a consistent answer set $T$ of $(P\setminus F)\cup E$ 
in which $G$ is true.  
By the definition of abductive programs, abducibles are assumed to appear 
in the head of no (non-factual) rule in $P$.  Thus, 
$a\in E$ and $b\in F$ imply $a\in T$ and $b\not\in T$, respectively. 
In this case, $UP^T$ contains facts $a\leftarrow$ and 
$\overline{b}\leftarrow$ which are respectively produced by 
$abd(a)$ and $abd(b)$. 
This implies that $UP^T$ contains a rule $\Sigma\leftarrow\Gamma$ with 
$\Sigma\subseteq {\cal L}_P$ 
iff $((P\setminus F)\cup E)^T$ has the same rule. 
Put $S=T\cup \{\,+a\mid a\in E\,\}\cup 
\{\,-b,\,\overline{b}\mid b\in F\,\}$. 
Then, $S$ is a consistent answer set of $UP\cup\{\,\leftarrow not\,G\,\}$, 
and $E^+=S\cap {\cal UA}^+$ and $F^-=S\cap {\cal UA}^-$. 

2. Suppose that $S$ is a consistent U-minimal answer set of 
$UP\cup \{\,\leftarrow not\,G\,\}$ 
such that $E^+=S\cap {\cal UA}^+$ and $F^-=S\cap {\cal UA}^-$. 
If the credulous explanation $(E,F)$ of $G$ is not minimal, 
there is a pair $(E',F')$ such that ($E'\subset E$ and $F'\subseteq F$) or 
($E'\subseteq E$ and $F'\subset F$), 
and $(P\setminus F')\cup E'$ has a consistent answer set $T'$ in which $G$ 
is true. 
Then, there is an answer set $S'$ of $UP\cup \{\,\leftarrow not\,G\,\}$ 
such that $T'=S'\cap {\cal L}_P$ and 
${E'}^+=S'\cap {\cal UA}^+$ and ${F'}^-=S'\cap {\cal UA}^-$ 
by the only-if part of~1. 
By $E'\cup F'\subset E\cup F$, ${E'}^+\cup {F'}^-\subset E^+\cup F^-$ holds. 
Thus, $S'\cap {\cal UA}\subset S\cap {\cal UA}$. 
This contradicts the assumption that $S$ is U-minimal. 
Conversely, when $(E,F)$ is a minimal credulous explanation of $G$, 
$UP\cup \{\,\leftarrow not\,G\,\}$ has a consistent answer set 
$S$ such that $E^+=S\cap {\cal UA}^+$ and $F^-=S\cap {\cal UA}^-$ (by 1). 
Suppose that $S$ is not U-minimal. Then, 
$UP\cup \{\,\leftarrow not\,G\,\}$ has a consistent U-minimal answer set 
$S'$ such that $S'\cap {\cal UA}\subset S\cap {\cal UA}$, 
${E'}^+=S'\cap {\cal UA}^+$, and ${F'}^-=S'\cap {\cal UA}^-$. 
In this case, there is a minimal credulous explanation $(E',F')$ of $G$ 
such that $E'\cup F'\subset E\cup F$ by the if-part of 2. 
This contradicts the fact that $(E,F)$ is minimal. 
Hence, the result holds. 
\end{proof}
\end{theorem}

\begin{theorem}[computing skeptical explanations by UP] \label{main-th1}
Let $\pf{P,{\cal A}}$ be an abductive program, $UP$ its update program, 
and $G$ a positive observation. Then, $G$ has a skeptical explanation 
$(E,F)$ iff $UP\,\cup\,\{\,\leftarrow not\,G\,\}$ has a consistent 
answer set $S$ 
such that $E^+=S\cap {\cal UA}^+$, $F^-=S\cap {\cal UA}^-$, and 
$(P\setminus F)\cup E\,\cup\,\{\,\leftarrow G\,\}$ is inconsistent. 
In particular, $(E,F)$ is a minimal skeptical explanation iff 
$S$ is U-minimal among those satisfying the above condition. 
\begin{proof}  
Suppose that $S$ is a consistent answer set of 
$UP\,\cup\,\{\,\leftarrow not\,G\,\}$ satisfying the condition 
that $E^+=S\cap {\cal UA}^+$, $F^-=S\cap {\cal UA}^-$, and 
$(P\setminus F)\cup E\,\cup\,\{\,\leftarrow G\,\}$ is inconsistent. 
Then, $(E,F)$ is a credulous explanation of $G$ (Theorem~\ref{vu-lem}), 
and also a skeptical explanation of $G$ (Proposition~\ref{pre-prop}). 
In particular, if $S$ is U-minimal among those satisfying the condition, 
$(E,F)$ becomes a minimal skeptical explanation by Theorem~\ref{vu-lem}. 

Conversely, suppose that $(E,F)$ is a skeptical explanation of $G$. 
By Proposition~\ref{pre-prop} and Theorem~\ref{vu-lem}, 
there is a consistent answer set $S$ of $UP\cup \{\,\leftarrow not\,G\,\}$ 
such that $E^+=S\cap {\cal UA}^+$, $F^-=S\cap {\cal UA}^-$, and 
$(P\setminus F)\cup E\,\cup\,\{\,\leftarrow G\,\}$ is inconsistent. 
Suppose that $(E,F)$ is a minimal skeptical explanation of $G$. 
To see that $S$ is U-minimal among those satisfying the condition, 
suppose that there is an answer set $S'$ which satisfies the condition 
and $S'\cap {\cal UA}\subset S\cap {\cal UA}$. 
Put ${E'}^+=S'\cap {\cal UA}^+$ and ${F'}^-=S'\cap {\cal UA}^-$. 
Then, ${E'}^+\cup {F'}^-\subset E^+\cup F^-$, 
thereby $E'\cup F'\subset E\cup F$. 
As $(E',F')$ is a skeptical explanation of $G$ 
by Proposition~\ref{pre-prop} and Theorem~\ref{vu-lem}, 
this contradicts the assumption that $(E,F)$ is minimal. 
\end{proof}
\end{theorem}

\begin{example} 
For the abductive program of Example~\ref{skpandcre-ex}, $UP$ becomes 
\begin{eqnarray*}
UP: &&  p\,; q\leftarrow a\,,\\
&& \neg q\leftarrow not\,b\,,\\
&& abd(a),\;\; abd(b),\\
&& +a\leftarrow a\,,\\
&& -b\leftarrow not\,b\,.
\end{eqnarray*}
For the positive observation $p$, the program 
$UP\cup\{\,\leftarrow not\,p\,\}$ has the answer set 
$S=\{\,p,a,\overline{b},+a,-b\,\}$ such that 
$E^+=\{+a\}$, $F^-=\{-b\}$, and 
$(P\setminus F)\cup E\,\cup\,\{\,\leftarrow p\,\}$ is inconsistent 
with $(E,F)=(\{a\},\{b\})$.  Since $S$ is also U-minimal satisfying 
this condition, $(\{a\},\{b\})$ is the minimal skeptical explanation of $p$. 
On the other hand, $UP\cup\{\,\leftarrow not\,p\,\}$ has another answer set 
$S'=\{\,p,a,b,+a\,\}$ such that $E^+=\{+a\}$ and $F^-=\emptyset$. 
However, $(P\setminus F)\cup E\,\cup\,\{\,\leftarrow p\,\}$ is consistent 
with $(E,F)=(\{a\},\{\})$, so that $(\{a\},\{\})$ 
is not a skeptical explanation (but a credulous one). 
\end{example}

The above results present that (minimal) explanations of 
extended abduction are computed by means of answer sets of 
an update program which is an EDP. 
In particular, when a program $P$ is an ELP (resp.\ NDP, NLP), 
explanations are computed by means of answer sets 
(resp.\ stable models) of the corresponding update program 
which is also an ELP (resp.\ NDP, NLP). 

For computing anti-explanations, we have the following results. 

\begin{lem}[converting anti-explanations to explanations]\label{anti-lem}
Let $\pf{P,{\cal A}}$ be an abductive program and $G$ a negative observation. 
Then, $(E,F)$ is a (minimal) credulous/skeptical anti-explanation of $G$ 
with respect to $\pf{P,{\cal A}}$ 
iff $(E,F)$ is a (minimal) credulous/skeptical explanation of a positive 
observation $G'$ with respect to the abductive program 
$\pf{P\cup\{\,G'\leftarrow not\,G\,\},{\cal A}}$, 
where $G'$ is a ground atom appearing nowhere in $P\cup {\cal A}$. 

In particular, 
$(E,F)$ is a (minimal) credulous anti-explanation of $G=\bot$ 
with respect to $\pf{P,{\cal A}}$ 
iff $(E,F)$ is a (minimal) credulous explanation of a positive 
observation $G'$ with respect to the abductive program 
$\pf{P\cup\{\,G'\leftarrow not\,\bot\,\},{\cal A}}$. 
\begin{proof}
Put $P'=P\,\cup\,\{\,G'\leftarrow not\,G\,\}$. 
Then, $G$ is not included in an answer set $S$ of a consistent program 
$(P\setminus F)\cup E$ 
iff $G'$ is included in an answer set $S\cup \{G'\}$ of a consistent program 
$(P'\setminus F)\cup E$. Hence, the result follows. 
In particular, 
when $G=\bot$,    
$(P\setminus F)\cup E$ is consistent  
iff $G'$ is included in a consistent answer set of $(P'\setminus F)\cup E$. 
\end{proof}
\end{lem}

\begin{theorem}[computing anti-explanations through UP]\label{main-th2}
Let $\pf{P,{\cal A}}$ be an abductive program, $UP$ its update program, 
and $G$ a negative observation. 
Also, let $G'$ be a ground atom appearing nowhere in $P\cup {\cal A}$, 
and $P'=P\cup\{\,G'\leftarrow not\,G\,\}$.  Then, 
\begin{enumerate}
\item $(E,F)$ is a (minimal) credulous anti-explanation of $G$ iff 
$UP\,\cup\,\{\,\leftarrow G\,\}$ has a consistent (U-minimal) answer set $S$ 
such that $E^+=S\cap {\cal UA}^+$ and $F^-=S\cap {\cal UA}^-$. 

\item $(E,F)$ is a skeptical anti-explanation of $G$ iff 
$UP\,\cup\,\{\,G'\leftarrow not\,G\,\}\,\cup\,\{\,\leftarrow not\,G'\,\}$ 
has a consistent answer set $S$ 
such that $E^+=S\cap {\cal UA}^+$, $F^-=S\cap {\cal UA}^-$, and 
$(P'\setminus F)\cup E\,\cup\,\{\,\leftarrow G'\,\}$ is inconsistent. 
In particular, $(E,F)$ is a minimal skeptical anti-explanation iff 
$S$ is U-minimal among those satisfying the above condition. 
\end{enumerate}

\begin{proof}\rm
1. Put $UP'=UP\,\cup\,\{\,G'\leftarrow not\,G\,\}$. 
Then, $(E,F)$ is a (minimal) credulous anti-explanation of $G$ 
with respect to $\pf{P,{\cal A}}$ \\ 
iff $(E,F)$ is a (minimal) credulous explanation of a positive observation 
$G'$ with respect to $\pf{P',{\cal A}}$ (Lemma~\ref{anti-lem})\\ 
iff $UP'\,\cup\,\{\,\leftarrow not\,G'\,\}$ has a consistent (U-minimal) answer set $S\cup \{\,G'\,\}$ such that 
$E^+=S\cap {\cal UA}^+$ and $F^-=S\cap {\cal UA}^-$ (by Theorem~\ref{vu-lem}). 
When $UP'\,\cup\,\{\,\leftarrow not\,G'\,\}$ has a consistent (U-minimal) answer set $S\cup \{\,G'\,\}$, $G$ is not included in $S$.  
So $UP'\,\cup\,\{\,\leftarrow not\,G'\,\}$ has a consistent (U-minimal) 
answer set $S\cup \{\,G'\,\}$ 
such that $E^+=S\cap {\cal UA}^+$ and $F^-=S\cap {\cal UA}^-$ iff 
$UP\cup \{\,\leftarrow G\,\}$ has a consistent (U-minimal) answer set $S$ 
such that $E^+=S\cap {\cal UA}^+$ and $F^-=S\cap {\cal UA}^-$. 

2. 
$(E,F)$ is a skeptical anti-explanation of $G$ 
with respect to $\pf{P,{\cal A}}$ \\ 
iff $(E,F)$ is a skeptical explanation of a positive observation 
$G'$ with respect to $\pf{P',{\cal A}}$ (Lemma~\ref{anti-lem})\\
iff $UP'\,\cup\,\{\,\leftarrow not\,G'\,\}$ has a consistent answer set $S$ 
such that $E^+=S\cap {\cal UA}^+$, $F^-=S\cap {\cal UA}^-$, and 
$(P'\setminus F)\cup E\,\cup\,\{\,\leftarrow G'\,\}$ is inconsistent. 
In particular, $(E,F)$ is a minimal skeptical anti-explanation iff 
$S$ is U-minimal among those satisfying the above condition 
(Theorem~\ref{main-th1}). 
\end{proof}
\end{theorem}

Suppose an abductive program $\pf{P,{\cal A}}$ such that $P$ is 
a normal logic program and ${\cal A}$ is a set of atoms. 
When $P$ is {\em locally stratified\/} in the sense of \cite{Prz88}, 
$P$ has at most one answer set (called a {\em perfect model\/}).
In this case, the above results are simplified as follows.%
\footnote{The result is generalized to the class of programs having 
at most one stable model.} 

\begin{cor}[computing (anti-)explanations in locally stratified NLPs]\label{strat-cor} 
Let $\pf{P,{\cal A}}$ be an abductive program in which 
$P$ is a locally stratified NLP and ${\cal A}$ is the set of abducible atoms. 
Also, let $UP$ be the update program of $\pf{P,{\cal A}}$ and 
$G$ a ground atom. 
Then, 
\begin{enumerate}
\item  A positive observation $G$ has a (minimal) explanation $(E,F)$ 
iff the program $UP\,\cup\,\{\,\leftarrow not\,G\,\}$
has a consistent (U-minimal) answer set $S$ such that 
$E^+=S\cap {\cal UA}^+$ and $F^-=S\cap {\cal UA}^-$.

\item  A negative observation $G$ has a (minimal) anti-explanation $(E,F)$ 
iff the program $UP\,\cup\,\{\,\leftarrow G\,\}$
has a consistent (U-minimal) answer set $S$ such that 
$E^+=S\cap {\cal UA}^+$ and $F^-=S\cap {\cal UA}^-$.

\end{enumerate}

\begin{proof}\rm
When $P$ is a locally stratified NLP, so is $(P\setminus F)\cup E$ 
because introducing/deleting facts to/from $P$ does not break the 
stratification structure. 
Then $(P\setminus F)\cup E$ has at most one answer set. 
In this case, credulous (anti-)explanations and 
skeptical (anti-)explanations coincide. 
Hence, the results hold by Theorems~\ref{vu-lem} and~\ref{main-th2}. 
\end{proof}
\end{cor}

The results of Theorems~\ref{vu-lem},~\ref{main-th1} and~\ref{main-th2} 
imply that 
any proof procedure for computing answer sets in EDPs is 
used for computing (anti-)explanations of extended abduction in EDPs. 
In particular, minimal (anti-)explanations are found by an additional 
mechanism of filtering U-minimal ones out of answer sets. 

\section{View updates through extended abduction} \label{sec:4}

In this section, we characterize the problem of view updates 
through extended abduction. 
We compute view updates by means of update programs in Section~\ref{sec:4.1}, 
and realize the task of integrity maintenance 
as a special case in Section~\ref{sec:4.2}. 

\subsection{View updates} \label{sec:4.1}

Suppose a knowledge base which contains variable rules and invariable rules. 
When there is a request for inserting/deleting a fact to/from the program, 
the update on the fact which is derived by invariable rules is translated 
into updates on variable rules/facts. 
This type of updates is called view updates. 

\begin{definition}[view updates]\label{def-vu}
Let $P$ be a program, $V$ the set of variable rules 
in the language of $P$, and $G$ a ground fact. 
Then, a program $P'$ accomplishes a {\em view update\/} for the 
{\em insertion\/} (resp.\ {\em deletion\/}) of $G$ to/from $P$ if 
\begin{enumerate}
\item $P'$ is consistent, 
\item $P'\models G$~~(resp.\, $P'\not\models G$), 
\item $P'\setminus V=P\setminus V$, 
\item there is no consistent program $P''$ such that 
$P''\models G$~~(resp.\, $P''\not\models G$),\\
$P''\setminus V=P\setminus V$, and 
$[(P\cap V)\sim (P''\cap V)]\subset [(P\cap V)\sim (P'\cap V)]$,\\
where $Q\sim R=(Q\setminus R)\cup (R\setminus Q)$. 
\end{enumerate}
\end{definition}

By the definition, the updated program $P'$ is a consistent program 
which minimally changes the variable part $V$ of $P$ to (un)imply $G$. 
Such a program $P'$ is obtained from $P$ by deleting some rules in 
$V\cap P$ and introducing some rules in $V\setminus P$. 
In particular, when $G\in V\setminus P$ (resp.\ $G\in V\cap P$), 
the insertion (resp.\ deletion) is done by directly introducing 
(resp.\ deleting) $G$ to/from $P$. 
We do not consider 
introducing rules in $V\cap P$ and deleting rules in $V\setminus P$, 
because introducing any rule which already exists in $P$ is redundant and 
deleting any rule which does not exist in $P$ is meaningless. 
With this assumption, the third condition $P'\setminus V=P\setminus V$ 
of Definition~\ref{def-vu} is equivalent to 
\[ P'=(P\setminus F)\cup E\;\;\mbox{ for }\;\; 
E\subseteq V\setminus P\;\;\mbox{ and }\;\; F\subseteq V\cap P\,.\] 

The view update problem is then naturally expressed 
by an abductive program $\pf{P,V}$, where the program $P$ 
represents a knowledge base and the abducibles $V$ represent   
variable rules. 

\begin{theorem}[view updates by extended abduction]\label{th-vu}
Let $P$ be a program and $V$ the set of variable rules in the 
language of $P$.  Given a ground literal $G$, 
$(P\setminus F)\cup E$ accomplishes a view update for 
inserting (resp.\ deleting) $G$ iff $(E,F)$ is a minimal skeptical 
explanation 
(resp.\ minimal credulous anti-explanation) of the positive observation 
(resp.\ negative observation) $G$ with respect to 
the abductive program $\pf{P,V}$. 
\begin{proof}
Suppose that $P'=(P\setminus F)\cup E$ accomplishes the insertion 
(resp.\ deletion) of $G$. 
Then, $P'$ is consistent and $P'\models G$ (resp.\ $P'\not\models G$). 
As $E\subseteq V\setminus P$ and $F\subseteq V\cap P$, 
$(E,F)$ is a skeptical explanation (resp.\ credulous anti-explanation) 
of $G$ with respect to $\pf{P,V}$. 
On the other hand, it holds that 
$(P'\cap V)\setminus (P\cap V)=E$ and  
$(P\cap V)\setminus (P'\cap V)=F$. 
Then, by the fourth condition of view updates, 
there is no $E'\subseteq V\setminus P$ 
nor $F'\subseteq V\cap P$ such that 
$(P\setminus F')\cup E'\models G$ 
(resp.\ $(P\setminus F')\cup E'\not\models G$) with 
a consistent $(P\setminus F')\cup E'$, and $E'\cup F'\subset E\cup F$. 
If $(P\setminus F')\cup E'$ is consistent and 
$(P\setminus F')\cup E'\models G$ 
(resp.\ $(P\setminus F')\cup E'\not\models G$), 
then $E'\subseteq E$ and $F'\subseteq F$ imply $E'=E$ and $F'=F$, 
because otherwise $E'\cup F'\subset E\cup F$. 
Thus, $(E,F)$ is a minimal skeptical explanation 
(resp.\ minimal credulous anti-explanation) of $G$ with respect to 
$\pf{P,V}$. 
The converse is obvious by the definition of minimal (anti-)explanations. 
\end{proof}
\end{theorem}

To realize view updates through update programs, we first transform 
the abductive program $\pf{P,V}$ to its normal form 
$\pf{P^{\rm n},V^{\rm n}}$ 
with abducible facts $V^{\rm n}$ as 
presented in Section~\ref{sec:2.2}. 
For $E\subseteq V$ and $F\subseteq V$, we define 
$n(E)^+=\{\,+a\mid a\in n(E)\,\}$ and $n(F)^-=\{\,-a\mid a\in n(F)\,\}$, 
where $n(\cdot)$ is the naming function introduced in Section~\ref{sec:2.2}. 
Then, the following results hold. 

\begin{theorem}[view insertion through UP] \label{vu-th1} 
Let $P$ be a program, $V$ the set of variable rules in the 
language of $P$, and $G$ a ground literal. 
Also, let $\pf{P^{\rm n},V^{\rm n}}$ be the normal form of the abductive 
program $\pf{P,V}$, 
and $UP$ the update program of $\pf{P^{\rm n},V^{\rm n}}$. 
Then, $(P\setminus F)\cup E$ accomplishes the insertion of $G$ iff 
\begin{enumerate}
\item $S$ is a consistent answer set of $UP\,\cup\,\{\,\leftarrow not\,G\,\}$ 
such that $n(E)^+=S\cap {\cal UA}^+$, $n(F)^-=S\cap {\cal UA}^-$, and 
$(P\setminus F)\cup E\,\cup\,\{\,\leftarrow G\,\}$ is inconsistent, and 
\item $S$ is U-minimal among those satisfying the condition 1. 
\end{enumerate}
\begin{proof} \rm
$(P\setminus F)\cup E$ accomplishes the insertion of $G$ \\
iff $(E,F)$ is a minimal skeptical explanation of $G$ with respect to 
$\pf{P,V}$ (Theorem~\ref{th-vu})\\
iff $(n(E),n(F))$ is a minimal skeptical explanation of $G$ with respect to 
$\pf{P^{\rm n},V^{\rm n}}$ (Proposition~\ref{naming-prop})\\ 
iff there exists a consistent U-minimal answer set $S$ 
of $UP\,\cup\,\{\,\leftarrow not\,G\,\}$ satisfying the conditions 
1 and 2 (Theorem~\ref{main-th1}). 
\end{proof}
\end{theorem}

\begin{theorem}[view deletion through UP]\label{vu-th2}
Let $P$ be a program, $V$ the set of variable rules in the 
language of $P$, and $G$ a ground literal. 
Also, let $\pf{P^{\rm n},V^{\rm n}}$ be the normal form of the 
abductive program $\pf{P,V}$, 
and let $UP$ be the update program of $\pf{P^{\rm n},V^{\rm n}}$. 
Then, $(P\setminus F)\cup E$ accomplishes the deletion of $G$ iff 
$UP\,\cup\,\{\,\leftarrow G\,\}$ has a consistent U-minimal answer set $S$ 
such that $n(E)^+=S\cap {\cal UA}^+$ and $n(F)^-=S\cap {\cal UA}^-$. 
\begin{proof}\rm
$(P\setminus F)\cup E$ accomplishes the deletion of $G$ \\
iff $(E,F)$ is a minimal credulous anti-explanation of $G$ with respect to $\pf{P,V}$ 
(Theorem~\ref{th-vu})\\
iff $(n(E),n(F))$ is a minimal credulous anti-explanation of $G$ with respect to 
$\pf{P^{\rm n},V^{\rm n}}$ (Proposition~\ref{naming-prop})\\ 
iff $S$ is a consistent U-minimal answer set 
of $UP\,\cup\,\{\,\leftarrow G\,\}$ 
such that $n(E)^+=S\cap {\cal UA}^+$ and $n(F)^-=S\cap {\cal UA}^-$ 
(Theorem~\ref{main-th2}). 
\end{proof}
\end{theorem}

\begin{example}\rm 
Let $P$ be the program and $V$ the set of variable rules such that 
\begin{eqnarray*}
P: && flies(x)\leftarrow bird(x), not\>ab(x),\\
   && ab(x)\leftarrow broken\mbox{-}wing(x),\\
   && bird(tweety)\leftarrow\mbox{},\\
   && bird(opus)\leftarrow\mbox{},\\
   && broken\mbox{-}wing(tweety)\leftarrow.\\
V: && broken\mbox{-}wing(x). 
\end{eqnarray*}
Then, $UP$ becomes 
\begin{eqnarray*}
UP: && flies(x)\leftarrow bird(x), not\>ab(x),\\
   && ab(x)\leftarrow broken\mbox{-}wing(x),\\
   && bird(tweety)\leftarrow\mbox{},\\
   && bird(opus)\leftarrow\mbox{},\\
   && abd(broken\mbox{-}wing(tweety)),\;\;\; 
      abd(broken\mbox{-}wing(opus)),\\
   && -broken\mbox{-}wing(tweety)\leftarrow not\; broken\mbox{-}wing(tweety)\,,\\
   && +broken\mbox{-}wing(opus)\leftarrow broken\mbox{-}wing(opus)\,. 
\end{eqnarray*}
To insert $flies(tweety)$, the U-minimal answer set of 
$UP\cup\{\,\leftarrow not\,flies(tweety)\,\}$ becomes 
$\{\,flies(tweety),\,flies(opus),\,bird(tweety),\,bird(opus),\,
\overline{broken\mbox{-}wing(tweety)},\newline 
\overline{broken\mbox{-}wing(opus)},\,
-broken\mbox{-}wing(tweety)\,\}$. 
Then, $(P\setminus F)\cup E$ accomplishes the insertion of $flies(tweety)$ 
with $(E,F)=(\emptyset,\{\,broken\mbox{-}wing(tweety)\,\})$. 

On the other hand, to remove $flies(opus)$, the U-minimal answer set of 
$UP\cup\{\,\leftarrow~flies(opus)\,\}$ becomes 
$\{\,bird(tweety),\,bird(opus),\,broken\mbox{-}wing(tweety),\newline 
broken\mbox{-}wing(opus),\,
ab(tweety),\,ab(opus),\,+broken\mbox{-}wing(opus)\,\}$. 
Then, $(P\setminus F)\cup E$ accomplishes the deletion of $flies(opus)$ 
with $(E,F)=(\{\,broken\mbox{-}wing(opus)\,\},\emptyset)$. 
\end{example} 

\subsection{Integrity maintenance} \label{sec:4.2} 

Integrity constraints are conditions that a knowledge base should satisfy 
through updates.  When integrity constraints are violated, 
variable rules/facts are modified to restore consistency. 
Such {\em integrity maintenance\/} 
is done as a special case of view updating. 

Let $I$ be the set of integrity constraints in a program $P$. 
Then, we say that $P$ {\em violates\/} integrity constraints from $I$ 
if $P\setminus I$ has no consistent answer set satisfying every rule in $I$. 
$P$ {\em satisfies\/} integrity constraints from $I$ if 
$P$ does not violate them.%
\footnote{This is the {\em consistency view\/} of integrity satisfaction \cite{SK88}.} 

\begin{definition}[integrity maintenance] \label{def-ic} 
Let $P$ be a program 
and $V$ the set of variable rules in the language of $P$. 
Also, let $I$ be the set of integrity constraints such that 
$I\subseteq P\setminus V$. 
Then, a program $P'$ {\em restores consistency\/} with respect to $I$ if 
\begin{enumerate}
\item $P'$ is consistent, 
\item $P'\setminus V=P\setminus V$, 
\item there is no consistent program $P''$ such that 
$P''\setminus V=P\setminus V$ and \\
$[(P\cap V)\sim (P''\cap V)]\subset [(P\cap V)\sim (P'\cap V)]$. 
\end{enumerate}
In particular, $P'=P$ if $P$ satisfies every integrity constraint in $I$.  
\end{definition}

The first condition implies that $P'$ satisfies every constraint in $I$. 
Note that 
by $I\subseteq P\setminus V$ every constraint in $I$ is invariable, so 
$I\subseteq P'$ holds by the second condition. 
The third condition requests the minimality of change. 
By the definition, integrity maintenance is defined as a special 
case of view deletion of Definition~\ref{def-vu} with $G=\bot$, i.e., 
$P'\not\models\bot$ is equivalent to the first condition. 
Then, the problem of integrity maintenance is 
characterized by an abductive program $\pf{P,V}$ 
and computed by its update program. 
The following results directly follow from 
Theorem~\ref{th-vu} and Theorem~\ref{vu-th2}. 

\begin{theorem}[integrity maintenance by extended abduction] \label{ic-prop} 
Let $P$ be a program, $I\subseteq P$ integrity constraints,  
and $V$ the set of variable rules in the language of $P$. 
Then, $(P\setminus F)\cup E$ restores consistency with respect to $I$ 
iff $(E,F)$ is a minimal credulous anti-explanation of the negative 
observation $G=\bot$ with respect to $\pf{P,V}$. 
\end{theorem}

\begin{theorem}[integrity maintenance through UP]\label{ic-th} 
Let $P$ be a program, $I\subseteq P$ integrity constraints,  
and $V$ the set of variable rules in the language of $P$. 
Also, let $\pf{P^{\rm n},V^{\rm n}}$ be the normal form of the abductive 
program $\pf{P,V}$, and $UP$ the update program of $\pf{P^{\rm n},V^{\rm n}}$. 
Then, $(P\setminus F)\cup E$ restores consistency with respect 
to $I$ iff $UP$ has a consistent U-minimal answer set $S$ 
such that $n(E)^+=S\cap {\cal UA}^+$ and $n(F)^-=S\cap {\cal UA}^-$. 
\end{theorem}

\begin{example} 
Let $\pf{P,V}$ be the abductive program such that 
\begin{eqnarray*}
P: && employee(john,35)\leftarrow,\\
&& manager(john)\leftarrow,\\
&& \leftarrow employee(x,y),\,manager(x),\,not\: talented(x),\, y<40\,.\\
V: && manager(x),\;\; talented(x). 
\end{eqnarray*}
The integrity constraint enforces the condition that 
any employee does not become a manager under the age 40 
unless he/she is talented.  
The $UP$ of this program becomes 
\begin{eqnarray*}
UP: 
&& employee(john,35)\leftarrow,\\
&& \leftarrow employee(x,y),\,manager(x),\,not\: talented(x),\,y<40\,.\\
&& abd(manager(x)),\;\; abd(talented(x)),\\
&& -manager(john)\leftarrow not\,manager(john),\\
&& +talented(x)\leftarrow talented(x), 
\end{eqnarray*}
which has two U-minimal answer sets: 
\begin{eqnarray*}
&&\{\,employee(john,35),\,\overline{manager(john)},\,
-manager(john),\,\newline \overline{talented(john)}\,\},\\
&&\{\,employee(john,35),\,manager(john),\,
+talented(john),\,talented(john)\,\}.
\end{eqnarray*}
That is, removing $manager(john)$ or inserting $talented(john)$ 
restores consistency with respect to the integrity constraint. 
\end{example}

\section{Theory updates} \label{sec:5}

In this section, we characterize the problem of theory updates 
through extended abduction. 
We first consider updating a knowledge base by a single rule 
in Section~\ref{sec:5.1}, 
then generalize the result to updating by a program in Section~\ref{sec:5.2}. 
Inconsistency removal is formalized as a special case of 
theory updates in Section~\ref{sec:5.3}. 

\subsection{Updates with a rule} \label{sec:5.1}

Suppose that an update request for inserting/deleting a rule is brought to 
a knowledge base in which every rule is variable. 
In this case, an update is done by directly inserting/deleting 
the rule to/from the program. 

\begin{definition}[updates with a rule]\label{df-ruleup} 
Let $P$ be a program and $R$ a rule such that $R\not\in P$. 
Then, $P'$ accomplishes the {\em insertion\/} of $R$ to $P$ if 
\begin{enumerate}
\item $P'$ is consistent, 
\item $\{R\}\subseteq P'\subseteq P\cup\{R\}$, 
\item there is no consistent program $P''$ such that 
$P'\subset P''\subseteq P\cup \{R\}$. 
\end{enumerate}
On the other hand, for a program $P$ and a rule $R$ such that $R\in P$, 
$P'$ accomplishes the {\em deletion\/} of $R$ from $P$ if 
\begin{enumerate}
\item $P'$ is consistent, 
\item $P'\subseteq P\setminus\{R\}$, 
\item there is no consistent program $P''$ such that 
$P'\subset P''\subseteq P\setminus\{R\}$. 
\end{enumerate}
\end{definition}

In the above definition, the second conditions present that the updated 
program $P'$ includes/excludes the rule $R$, and 
the third conditions present that $P'$ minimally changes the original program 
$P$ by inserting/deleting $R$ to/from $P$. 

We first show that the problem of deleting a rule from a program 
in Definition~\ref{df-ruleup} is converted to the problem of 
inserting a rule to a program. 

\begin{prop}[converting deletion of a rule to insertion of a rule]\label{convert-prop} 
Let $P$ be a program and $R$ a rule in $P$. 
Then, there is a program $P'$ which accomplishes the deletion of $R$ from $P$ 
iff there is a program $PR'$ which accomplishes the insertion of the 
rule $\leftarrow\gamma_R$ to the program 
$PR=(P\setminus\{R\})\cup\{\, \Sigma\leftarrow\Gamma,\gamma_R,\;\;\gamma_R\leftarrow\,\}$ where $R=(\Sigma\leftarrow\Gamma)$. 
\begin{proof}
Suppose that $P'$ accomplishes the deletion of $R$ from $P$. 
Put $PR'=P'\,\cup\,\{\,\Sigma\leftarrow\Gamma,\gamma_R,\,\;\; \leftarrow~\gamma_R\,\}$. 
Then, by $P'\subseteq P\setminus\{R\}$, 
$PR'\subseteq (P\setminus\{R\})\,\cup\,\{\,\Sigma\leftarrow\Gamma,\gamma_R,\,\;\; \leftarrow~\gamma_R\,\}$ holds, thereby 
$\{\,\leftarrow\gamma_R\,\}\subseteq PR'\subseteq PR\,\cup\,\{\,\leftarrow\gamma_R\,\}$. 
As $P'$ is consistent, $PR'$ is consistent. 
Assume that there is a consistent program $PR''$ such that 
$PR'\subset PR''\subseteq PR\cup \{\,\leftarrow\gamma_R\,\}$. 
Put $P''=PR''\setminus \{\,\Sigma\leftarrow \Gamma,\gamma_R,\; \leftarrow\gamma_R\,\}$. 
Then, $PR''\subseteq PR\cup \{\,\leftarrow\gamma_R\,\}$ implies 
$PR''\subseteq (P\setminus\{R\})\,\cup\,
\{\,\Sigma\leftarrow \Gamma,\gamma_R,\; \gamma_R\leftarrow\,\}
\,\cup\,\{\,\leftarrow\gamma_R\,\}$, thereby 
$P''\subseteq P\setminus\{R\}$. 
On the other hand, $PR'\subset PR''$ implies 
$P'\,\cup\,\{\,\Sigma\leftarrow\Gamma,\gamma_R,\,\;\; \leftarrow~\gamma_R\,\}\subset PR''$, thereby $P'\subset PR''\setminus\{\,\Sigma\leftarrow\Gamma,\gamma_R,\,\;\; \leftarrow~\gamma_R\,\}$.  Then, $P'\subset P''$. 
Thus, $P'\subset P''\subseteq P\setminus\{R\}$ holds, 
which contradicts the fact that there is no such $P''$. 
Hence, $PR'$ accomplishes the insertion of $\leftarrow\gamma_R$ to $PR$. 

Conversely, suppose that 
$PR'$ accomplishes the insertion of $\leftarrow\gamma_R$ to $PR$. 
Put $P'=PR'\setminus \{\,\Sigma\leftarrow\Gamma,\gamma_R,\;\;\leftarrow\gamma_R\,\}$. 
Then, by $PR'\subseteq PR\cup\{\,\leftarrow\gamma_R\,\}$, 
$P'\subseteq P\setminus\{R\}$ holds. 
As $PR'$ is consistent, $P'$ is consistent. 
Assume that there is a consistent program $P''$ such that 
$P'\subset P''\subseteq P\setminus\{R\}$. 
Put $PR''=P''\cup\{\,\Sigma\leftarrow\Gamma,\gamma_R,\; \leftarrow\gamma_R\,\}$. 
Then, by $P'\cup\{\,\Sigma\leftarrow\Gamma,\gamma_R,\; \leftarrow\gamma_R\,\}=PR'$ and $P\setminus\{R\}\cup 
\{\,\Sigma\leftarrow\Gamma,\gamma_R,\; \leftarrow\gamma_R\,\}
\subseteq PR\cup\{\,\leftarrow\gamma_R\,\}$, it holds that 
$PR'\subset PR''\subseteq PR\cup\{\,\leftarrow\gamma_R\,\}$, 
which contradicts the fact that there is no such $PR''$. 
Hence, $P'$ accomplishes the deletion of $R$ from $P$. 
\end{proof}
\end{prop}

By Proposition~\ref{convert-prop}, for updating a program with a rule, 
it is enough to consider the problem of inserting a rule to a program. 
We study the problem in a more general setting in the next subsection. 

\subsection{Updates with programs} \label{sec:5.2} 

This section considers an update which updates a program with another 
program. 
Given a program $P$ which represents the current knowledge base 
and another program $Q$ which represents new information, 
a theory update is defined to satisfy the following conditions. 

\begin{definition}[theory updates]\label{df-th-up} 
Given programs $P$ and $Q$, 
$P'$ accomplishes a {\em theory update\/} of $P$ by $Q$ if 
\begin{enumerate}
\item $P'$ is consistent, 
\item $Q\subseteq P'\subseteq P\cup Q$, 
\item there is no consistent program $P''$ such that 
$P'\subset P''\subseteq P\cup Q$. 
\end{enumerate}
\end{definition}

By the definition, the updated program $P'$ is defined as 
the union of the new information $Q$ and 
a maximal subset of the original program $P$ which is consistent with $Q$. 
The first condition implies that 
new information $Q$ should be consistent, namely, 
updating with inconsistent information makes no sense. 
With this definition, inserting a rule to a theory of 
Definition~\ref{df-ruleup} 
is captured as a special case of a theory update of Definition~\ref{df-th-up} 
in which a new program $Q$ is given as a single rule. 
In contrast to this, it is considered a theory update which is defined 
as the removal of $Q$ from $P$ like $P'\subseteq P\setminus Q$. 
For such updates, the transformation of Proposition~\ref{convert-prop} is 
applied for each rule in $Q$.  Then the problem of removing $Q$ is converted 
to the problem of introducing corresponding 
rules as in Definition~\ref{df-th-up}. 

To realize theory updates, an abductive framework is used for specifying 
priorities between the current knowledge and the new knowledge.  
Consider the abductive program $\pf{P\,\cup\, Q,P\,\setminus\, Q}$, 
where a program is given as $P\,\cup\,Q$ and 
any rule in the original program $P$ other than 
the new information $Q$ is specified as variable abducible rules. 

\begin{theorem}[theory updates by extended abduction]\label{th-tu} 
Let $P$ and $Q$ be programs. 
Then, $P'$ accomplishes a theory update of $P$ by $Q$ iff 
$P'=(P\,\cup\,Q)\,\setminus\,F$ where 
$(\emptyset,F)$ is a minimal credulous anti-explanation of the negative 
observation $G=\bot$ 
with respect to the abductive program $\pf{P\,\cup\,Q,P\,\setminus\,Q}$. 
\begin{proof}\rm 
$P'$ accomplishes a theory update of $P$ by $Q$\\
iff $P'=(P\cup Q)\setminus F$ where 
$F$ is a minimal set such that $F\subseteq P\setminus Q$ and 
$(P\cup Q)\setminus F\not\models \bot$\\
iff $P'=(P\cup Q)\setminus F$ where 
$(\emptyset,F)$ is a minimal credulous anti-explanation of the negative observation 
$G=\bot$ with respect to $\pf{P\cup Q,P\setminus Q}$. 
\end{proof}
\end{theorem}

The abductive program $\pf{P\cup Q,P\setminus Q}$ is transformed to 
the normal form $\pf{(P\cup Q)^{\rm n},(P\setminus Q)^{\rm n}}$ 
where $(P\setminus Q)^{\rm n}$ consists of abducible facts 
(Section~\ref{sec:2.2}). 
Then, a minimal credulous anti-explanation of $G=\bot$ is computed by 
a consistent U-minimal answer set of the update program of 
$\pf{(P\cup Q)^{\rm n},(P\setminus Q)^{\rm n}}$. 
Note that in $\pf{(P\cup Q)^{\rm n},(P\setminus Q)^{\rm n}}$ 
it holds that $(P\setminus Q)^{\rm n}\setminus (P\cup Q)^{\rm n}=\emptyset$, 
so $UP$ contains no rule of the form $+a\leftarrow a$ of Definition~\ref{def-ur}(2). 

\begin{theorem}[theory updates through UP]\label{main-th3} 
Let $P$ and $Q$ be programs, and $UP$ the update program of 
the abductive program $\pf{(P\cup Q)^{\rm n},(P\setminus Q)^{\rm n}}$. 
Then, $(P\cup Q)\setminus F$ accomplishes a theory update of $P$ by $Q$ 
iff $UP$ has a consistent U-minimal answer set $S$ 
such that $n(F^-)=S\cap {\cal UA}^-$. 
\begin{proof}\rm 
$(P\cup Q)\setminus F$ accomplishes a theory update of $P$ by $Q$ \\
iff $(\emptyset,F)$ is a minimal credulous anti-explanation of the negative observation 
$G=\bot$ with respect to $\pf{P\cup Q,P\setminus Q}$ 
(Theorem~\ref{th-tu})\\
iff $(\emptyset,n(F))$ is a minimal credulous anti-explanation of $G=\bot$ 
with respect to $\pf{(P\cup Q)^{\rm n},(P\setminus Q)^{\rm n}}$ (Proposition~\ref{naming-prop})\\ 
iff $UP\cup\,\{\,\leftarrow\bot\,\}$ has a consistent U-minimal answer set $S$ 
such that $n(F^-)=S\cap {\cal UA}^-$ (Theorem~\ref{main-th2}).\\
iff $UP$ has a consistent U-minimal answer set $S$ 
such that $n(F^-)=S\cap {\cal UA}^-$. 
\end{proof}
\end{theorem}

\begin{example}[Alferes \emph{et al.}\shortcite{AL00}]\label{program-ex}
Given the current knowledge base 
\begin{eqnarray*}
P_1: && sleep\leftarrow not\,\mbox{\it tv\_on}\,,\\
   &&  watch\_tv\leftarrow  tv\_on\,,\\
   &&  tv\_on\leftarrow\,,
\end{eqnarray*}
consider updating $P_1$ with%
\footnote{In \cite{AL00} the rule ``$\leftarrow power\_failure,\,tv\_on$'' 
is given as ``$not\, \mbox{\it tv\_on} \leftarrow  power\_failure$''. 
These two rules are semantically equivalent 
under the answer set semantics \cite{IS98}.\label{fn-glp}}
\begin{eqnarray*}
P_2: && power\_failure\leftarrow\,,\\
&& \leftarrow  power\_failure,\,tv\_on\,.
\end{eqnarray*}
The situation is expressed by the abductive program 
$\pf{P_1\cup P_2,P_1\setminus P_2}$. 
The update program $UP$ of $\pf{(P_1\cup P_2)^{\rm n},(P_1\setminus P_2)^{\rm n}}$ 
then becomes 
\begin{eqnarray*}
UP: 
    && power\_failure\leftarrow\,,\\
    && \leftarrow  power\_failure,\,tv\_on\,,\\
    && sleep\leftarrow not\,\mbox{\it tv\_on},\,\gamma_1\,,\\
    && watch\_tv\leftarrow tv\_on,\,\gamma_2\,,\\
    && abd(tv\_on),\;\;\; abd(\gamma_1),\;\;\;  abd(\gamma_2),\\
    && -tv\_on\leftarrow not\, \mbox{\it tv\_on}\,,\\
    &&  -\gamma_1\leftarrow not\,\gamma_1\,,\\
    &&  -\gamma_2\leftarrow not\,\gamma_2\,,
\end{eqnarray*}
where $\gamma_1$ and $\gamma_2$ are the names of the abducible rules in 
$P_1\setminus P_2$. 
Then, $UP$ has the unique U-minimal answer set 
$\{\, power\_failure,\, sleep,\, \overline{tv\_on},\, 
-tv\_on,\, \gamma_1,\, \gamma_2\,\}$, 
which represents the deletion of the fact $tv\_on$ from $P_1\cup P_2$. 
As a result, the theory update of $P_1$ by $P_2$ becomes 
\begin{eqnarray*}
P_3: && sleep\leftarrow not\,\mbox{\it tv\_on}\,,\\
   &&  watch\_tv\leftarrow  tv\_on\,,\\
   && power\_failure\leftarrow\,,\\
   && \leftarrow  power\_failure,\,tv\_on\,.
\end{eqnarray*}

Next, suppose that another update 
\[ P_4:\;  \neg\,power\_failure\leftarrow \]
is given to $P_3$ which states that power is back again. 
The situation is expressed by the abductive program 
$\pf{P_3\cup P_4,P_3\setminus P_4}$, and the update program of 
$\pf{(P_3\cup P_4)^{\rm n},(P_3\setminus P_4)^{\rm n}}$ becomes 
\begin{eqnarray*}
UP: && \neg\,power\_failure\leftarrow\,,\\
    && sleep\leftarrow not\,\mbox{\it tv\_on}\,,\gamma_1\,,\\
    && watch\_tv\leftarrow tv\_on\,,\gamma_2\,,\\
   && \leftarrow  power\_failure,\,tv\_on\,,\gamma_3\,,\\
    && abd(power\_failure),\;\;\; abd(\gamma_1),\; abd(\gamma_2),\; 
abd(\gamma_3),\\
    && -power\_failure\leftarrow not\,\mbox{\it power\_failure}\,,\\
    && -\gamma_1\leftarrow not\,\gamma_1,\;\; 
       -\gamma_2\leftarrow not\,\gamma_2,\;\; 
       -\gamma_3\leftarrow not\,\gamma_3\,.
\end{eqnarray*}
Then, $UP$ has the unique U-minimal answer set 
$\{\, \neg\,power\_failure,\, sleep,\, \gamma_1,\, \gamma_2,\,\newline
\gamma_3,\, \overline{power\_failure},\, -power\_failure\,\}$, which 
implies that the result of the update is 
$(P_3\cup P_4)\setminus \{\,power\_failure\leftarrow\,\}$. 
\end{example}

Generally, there are several solutions for updating a program $P$ by $Q$. 
For example, let 
$P=\{\,p\leftarrow q,\;\; q\leftarrow\,\}$ and $Q=\{\,\neg p\leftarrow\,\}$. 
Then, there are two solutions of updating $P$ by $Q$; 
removing either $p\leftarrow q$ or $q\leftarrow$ from $P$. 
Every answer set which results from multiple solutions is 
expressed by a single program as follows. 

Suppose updating $P$ by $Q$.  Then, define the program 
\[\Pi=Q\cup \{\,\Sigma\leftarrow \Gamma,\,\gamma_R,\;\; abd(\gamma_R)\,\mid 
R=(\Sigma\leftarrow \Gamma)\in P\,\}\,.\]
Let $\Delta=\{\,\gamma_R\,\mid\,\gamma_R\;\mbox{appears in}\;\Pi\,\}$. 
A consistent answer set $S$ of $\Pi$ is called $\Delta$-{\em maximal\/} if 
$S$ is an answer set of $\Pi$ such that 
$T\cap\Delta\subseteq S\cap\Delta$ for any answer set $T$ of $\Pi$. 
Let ${\cal L}_{P\cup Q}$ be the set of all ground literals in 
the language of the program $P\cup Q$. 
Then the following result holds. 

\begin{theorem}[representing multiple solutions in a single program]\label{single-th}
Let $P$ and $Q$ be programs and $P'$ a result of a theory update of $P$ by $Q$. 
Also, let $\Pi$ be a program defined as above. 
Then, for any answer set $S$ of $P'$, 
there is a $\Delta$-maximal answer set $T$ of $\Pi$ 
such that $S=T\cap {\cal L}_{P\cup Q}$. 
Conversely, for any $\Delta$-maximal answer set $T$ of $\Pi$, 
there is a program $P'$ which has an answer set $S$ such that 
$S=T\cap {\cal L}_{P\cup Q}$. 
\begin{proof}
When $P'$ accomplishes a theory update of $P$ by $Q$, 
$P'=Q\cup P''$ where $P''$ is a maximal subset of $P$ 
such that $Q\cup P''$ is consistent.  Consider the program 
$\Pi'=Q\cup \{\,\Sigma\leftarrow \Gamma,\,\gamma_R,\; \gamma_R\leftarrow\,\mid R=(\Sigma\leftarrow \Gamma)\in P''\,\}$. 
Then, for any answer set $S$ of $P'$, there is an answer set $T'$ of $\Pi'$ 
such that $S=T'\cap {\cal L}_{P\cup Q}$. 
In this case, there is an answer set $T$ of $\Pi$  
such that $T=T'\cup \{\,\overline{\gamma_R}\,\mid\,\gamma_R\in\Delta\setminus T'\,\}$. 
Suppose that $T$ is not $\Delta$-maximal. 
Then, there is an answer set $T''$ of $\Pi$ such that 
$T'\cap\Delta\subset T''\cap\Delta$. 
In this case, there is a consistent answer set $S'$ of $Q\cup P'''$ 
such that $S'=T''\cap {\cal L}_{P\cup Q}$ and $P''\subset P'''\subseteq P$. 
This contradicts the assumption that $P''$ is a maximal subset of $P$ 
such that $Q\cup P''$ is consistent. 
Hence, $T$ is a $\Delta$-maximal answer set of $\Pi$. 
The converse is shown in a similar manner. 
\end{proof}
\end{theorem}

\begin{example}
In the above example, the program $\Pi$ becomes 
\begin{eqnarray*}
\Pi:&& \neg p\leftarrow,\\
&& p\leftarrow q,\,\gamma_1,\\
&& q\leftarrow \gamma_2,\\
&& abd(\gamma_1),\;\; abd(\gamma_2).
\end{eqnarray*}
Then, the $\Delta$-maximal answer sets of $\Pi$ are 
$\{\,\neg p,\,\gamma_1,\,\overline{\gamma_2}\,\}$ and 
$\{\,\neg p,\,q,\,\overline{\gamma_1},\,\gamma_2\,\}$, 
which correspond to the answer sets of the updated programs 
$\{\,p\leftarrow q,\;\; \neg p\leftarrow\,\}$ and 
$\{\,q\leftarrow,\;\; \neg p\leftarrow\,\}$, respectively. 
\end{example}

\subsection{Inconsistency removal}  \label{sec:5.3} 

A knowledge base may become inconsistent 
by the presence of contradictory information. 
In this situation, a knowledge base must be updated to restore 
consistency by detecting the source of inconsistency in the program. 
Such an inconsistency removal is defined as follows. 

\begin{definition}[inconsistency removal]
Let $P$ be a program. 
Then, a  program $P'$ accomplishes an {\em inconsistency removal\/} of $P$ if 
\begin{enumerate}
\item $P'$ is consistent, 
\item $P'\subseteq P$, 
\item there is no consistent program $P''$ such that $P'\subset P''\subseteq P$. 
\end{enumerate}
In particular, $P'=P$ if $P$ is consistent. 
\end{definition}  

By the definition, inconsistency removal is captured as a 
special case of theory updates where $P$ is possibly inconsistent and 
$Q$ is empty in Definition~\ref{df-th-up}.  
Then, by putting $Q=\emptyset$ in $\pf{P\cup Q,P\setminus Q}$, 
inconsistency removal is characterized by the abductive program $\pf{P,P}$. 
The next theorem directly follows from Theorem~\ref{th-tu}. 

\begin{theorem}[inconsistency removal by extended abduction]\label{th-ir}
Let $P$ be a program. 
Then, $P'$ accomplishes an inconsistency removal of $P$ iff 
$P'=P\setminus F$ where $(\emptyset,F)$ is a minimal credulous anti-explanation of 
the negative observation $G=\bot$ 
with respect to the abductive program $\pf{P,P}$. 
\end{theorem}

In an EDP, inconsistency arises when 
a program $P$ has the contradictory answer set ${\cal L}_P$ or 
$P$ has no answer set.  
An abductive program $\pf{P,P}$ can remove these different types of 
inconsistencies. 

\begin{example} 
Let $P=\{\,p\leftarrow not\,p,\;\; q\leftarrow\,\}$ which has no answer set. 
Then, $G=\bot$ has the minimal credulous (and also skeptical) anti-explanation 
$(E,F)=(\emptyset,\{\,p\leftarrow not\,p\,\})$ with respect to $\pf{P,P}$.  As a result, 
$P'=\{\,q\leftarrow\,\}$ accomplishes an inconsistency removal of $P$. 
\end{example}

The following result holds by Theorem~\ref{main-th3}.

\begin{theorem}[inconsistency removal through UP]
Let $P$ be a program and $UP$ 
the update program of the abductive program $\pf{P^{\rm n},P^{\rm n}}$ 
which is a normal form of $\pf{P,P}$. 
Then, $P\setminus F$ accomplishes an inconsistency removal of $P$ 
iff $UP$ has a consistent U-minimal answer set $S$ 
such that $n(F^-)=S\cap {\cal UA}^-$. 
\end{theorem}

\begin{example}  \label{nixon-diamond}
Let $P$ be the program 
\begin{eqnarray*}
&& pacifist\leftarrow quaker\,,\\
&& \neg pacifist\leftarrow republican\,,\\
&& quaker\leftarrow\,,\;\;\; republican\leftarrow\,, 
\end{eqnarray*}
which has the answer set ${\cal L}_P$. 
Consider the update program $UP$ of the abductive program $\pf{P^{\rm n},P^{\rm n}}$: 
\begin{eqnarray*}
UP: && pacifist\leftarrow quaker,\gamma_1\,,\\
&& \neg pacifist\leftarrow republican,\gamma_2\,,\\
&& abd(\gamma_1),\;\; abd(\gamma_2),\;\; abd(quaker),\;\; abd(republican),\\
&& -\gamma_1\leftarrow not\,\gamma_1,\\
&& -\gamma_2\leftarrow not\,\gamma_2,\\
&& -quaker\leftarrow not\,\mbox{\it quaker}\,,\\
&& -republican\leftarrow not\,\mbox{\it republican}\,. 
\end{eqnarray*}
Then, $UP$ has four U-minimal answer sets: 
\begin{eqnarray*}
&&\{\,quaker,\,republican,\, pacifist,\,\gamma_1,\,\overline{\gamma_2}, -\gamma_2\,\},\\
&&\{\,quaker,\, republican,\,\neg pacifist,\,\overline{\gamma_1},\,\gamma_2,\,-\gamma_1\,\},\\
&&\{\,\overline{quaker},\, republican,\,\gamma_1,\, \gamma_2,\, \neg pacifist,\,-quaker\,\},\\
&&\{\,quaker,\, \overline{republican},\,pacifist,\,\gamma_1,\,\gamma_2,\, -republican\,\},
\end{eqnarray*}
which represent that deletion of one of the rules (or facts) from $P$ 
makes the program consistent.  
\end{example}

The multiplicity of possible solutions as in the above example is expressed 
by a single program using the technique of Theorem~\ref{single-th}. 
On the other hand, if one wants to restrict the set of rules to be removed, 
it is done by considering an abductive program $\pf{P,P'}$ with 
$P'\subseteq P$.  In this case, any rule in $P'$ is subject to change 
to recover consistency. 

\section{Computational complexity} \label{sec:6}

In this section, we compare the computational complexity of 
different types of updates. 
Throughout the section, we consider propositional abductive programs, i.e., 
an abductive program 
$\pf{P,{\cal A}}$ where $P$ is a finite EDP containing 
no variable and ${\cal A}$ is a finite set of ground literals. 
An observation $G$ is a ground literal. 
We also assume an abductive program $\pf{P,{\cal A}}$ where 
${\cal A}$ consists of abducible facts.  An abductive program with 
abducible rules is transformed to an abductive program with 
abducible facts by considering its normal form (see Section~\ref{sec:2.2}). 

We first investigate the complexity of extended abduction. 
The decision problems considered here are analogous to those of \cite{EGL97}, 
that is, given an abductive program $\pf{P,{\cal A}}$ and a 
positive/negative observation $G$: 
\begin{description}
\item[Existence:] Does $G$ have an (anti-)explanation with respect to 
$\pf{P,{\cal A}}$? 
\item[Relevance:] Is a given abducible $A\in {\cal A}$ included in some 
(anti-)explanation $(E,F)$ of $G$ (i.e.,  $A\in E\cup F$)? 
\item[Necessity:] Is a given abducible $A\in {\cal A}$ included in every 
(anti-)explanation of $G$? 
\end{description}
Since the existence of (anti-)explanations implies 
the existence of minimal (anti-)explanations, deciding the existence of 
a minimal (anti-)explanation is as hard as deciding the 
existence of an arbitrary one. 
Similarly, considering minimal (anti-)explanations instead of arbitrary 
ones brings the same result in the necessity problem. 
By contrast, the relevance problem has different complexity results 
between arbitrary and minimal (anti-)explanations in general.  

To analyze the complexity of each problem, 
we first introduce a transformation from 
extended abduction to normal abduction based on the one in \cite{Ino00}.%
\footnote{Recall that by normal abduction we mean abduction which 
explains a positive observation only by introducing hypotheses.} 
This transformation enables us to use the complexity results of 
normal abduction. 

Suppose an abductive program $\pf{P,{\cal A}}$ where ${\cal A}$ consists of 
abducible facts. 
We define an abductive program $\pf{P',{\cal A}'}$ such that 
\begin{eqnarray*}
P'&=& (P\setminus {\cal A})\;\cup\; \{\,A\leftarrow not\,A'\,\mid\,A\in {\cal A}\cap P\,\},\\
{\cal A}'&=& ({\cal A}\setminus P)\,\cup\,\{\,A'\,\mid\,A\in {\cal A}\cap P\,\}, 
\end{eqnarray*}
where $A'$'s are ground literals associated with each $A$ and appear nowhere 
in $P\cup {\cal A}$. 
In the abductive program $\pf{P',{\cal A}'}$, any abducible in 
${\cal A}\cap P$ is made non-abducible 
and a new abducible $A'$ is introduced for each $A\in {\cal A}\cap P$. 
With this setting, the removal of $A\in {\cal A}\cap P$ from $P$ is 
achieved by the introduction of $A'\in {\cal A'}$ to $P'$ 
by the rule $A\leftarrow not\,A'$. 
The next proposition is due to \cite{Ino00}.%
\footnote{
The proposition is given in a more general setting in \cite{Ino00}, 
but the definition of (anti-)explanations in \cite{Ino00} is a bit 
different from the one in this paper.}

\begin{prop}[transformation from extended abduction to normal abduction]\label{trans-xabd}
Let $\pf{P,{\cal A}}$ be an abductive program and $G$ a ground literal. 
\begin{enumerate}
\item A positive observation $G$ has a (minimal) credulous/skeptical 
explanation 
$(E,F)$ with respect to $\pf{P,{\cal A}}$ under extended abduction 
iff $G$ has a (minimal) credulous/skeptical explanation 
$H=E\cup \{\,A'\,\mid\,A\in F\,\}$ 
with respect to $\pf{P',{\cal A}'}$ under normal abduction. 

\item A negative observation $G$ has a (minimal) credulous/skeptical 
anti-explanation 
$(E,F)$ with respect to $\pf{P,{\cal A}}$ under extended abduction 
iff $G'$ has a (minimal) credulous/skeptical explanation 
$H=E\cup \{\,A'\,\mid\,A\in F\,\}$ with respect to 
$\pf{P'\cup\{\,G'\leftarrow not\,G\,\},{\cal A}'}$ under normal abduction, 
where $G'$ is a ground atom appearing nowhere in $P\cup {\cal A}$. 
\end{enumerate}

\begin{proof}
1. By the definition, 
an abducible $A\in {\cal A}\cap P$ is not in $P\setminus F$ iff 
$A'\in {\cal A}'$ is in $P'\cup \{\,A'\,\mid\,A\in F\,\}$. 
Then, $(P\setminus F)\cup E$ has an answer set $S$ iff 
$P'\cup H$ has an answer set $S\cup\{\,A'\,\mid\,A\in F\,\}$. 
Hence, $G$ has a credulous/skeptical explanation 
$(E,F)$ with respect to $\pf{P,{\cal A}}$ under extended abduction 
iff $G$ has a credulous/skeptical explanation 
$H=E\cup \{\,A'\,\mid\,A\in F\,\}$ 
with respect to $\pf{P',{\cal A}'}$ under normal abduction. 
In particular, $(E,F)$ is minimal iff $E\cup F$ is minimal iff $H$ is minimal. 

2. By Lemma~\ref{anti-lem}, 
$G$ has a (minimal) credulous/skeptical anti-explanation $(E,F)$ with 
respect to $\pf{P,{\cal A}}$ under extended abduction 
iff a positive observation $G'$ has a (minimal) credulous/skeptical 
explanation $(E,F)$ 
with respect to $\pf{P\,\cup\,\{\,G'\leftarrow not\,G\,\},{\cal A}}$ under 
extended abduction. 
Then, the result holds by the part 1 of this proposition. 
\end{proof}
\end{prop}

Thus, extended abduction is efficiently converted into normal abduction. 
On the other hand, normal abduction is captured as a special case of 
extended abduction. That is, given an abductive program 
$\pf{P,{\cal A}}$ and a positive observation $G$, 
$G$ has a (minimal) credulous/skeptical 
explanation $E$ with respect to $\pf{P,{\cal A}}$ 
under normal abduction iff $G$ has a 
(minimal) credulous/skeptical explanation $(E,\emptyset)$ 
with respect to $\pf{P,{\cal A}}$ under extended abduction. 

We use these results for assessing the complexity of extended abduction. 

\begin{prop}[complexity results for normal abduction \cite{EGL97}]\label{complex-lem}
Given a propositional abductive program $\pf{P,{\cal A}}$ 
and a ground positive observation $G$: 
\begin{description}
\item[](a) Deciding if $G$ has a credulous/skeptical explanation  
is $\Sigma_2^P$-complete/$\Sigma_3^P$-complete. 
\item[](b) Deciding if an abducible $A\in {\cal A}$ is relevant to 
some credulous/skeptical explanation 
(resp.\ some {\em minimal\/} credulous/skeptical explanation) of $G$ is 
$\Sigma_2^P$-complete/$\Sigma_3^P$-complete 
(resp.\ $\Sigma_3^P$-complete/$\Sigma_4^P$-complete). 
\item[](c) Deciding if an abducible $A\in {\cal A}$ is necessary for every 
(minimal) credulous/skeptical explanation of $G$ is 
$\Pi_2^P$-complete/$\Pi_3^P$-complete. 
\end{description}
In particular, when $P$ contains no disjunctive rules (i.e., $P$ is an ELP), 
the complexity of each problem decreases by one level in the polynomial 
hierarchy.%
\footnote{In \cite{EGL97} the results are reported for 
normal logic/disjunctive programs, but the same results hold for 
extended logic/disjunctive programs.}
\end{prop}

\begin{theorem}[complexity results for extended abduction]\label{complexity-th}
Let $\pf{P,{\cal A}}$ be a propositional abductive program. 
\begin{enumerate}
\item Given a ground positive observation $G$: 
\begin{enumerate}
\item Deciding if $G$ has a credulous/skeptical explanation 
is $\Sigma_2^P$-complete/$\Sigma_3^P$-complete. 
\item Deciding if an abducible $A\in {\cal A}$ is relevant to 
some credulous/skeptical explanation 
(resp.\ some {\em minimal\/} credulous/skeptical explanation) of $G$ is 
$\Sigma_2^P$-complete/$\Sigma_3^P$-complete 
(resp.\ $\Sigma_3^P$-complete/$\Sigma_4^P$-complete). 
\item Deciding if an abducible $A\in {\cal A}$ is necessary for every 
(minimal) credulous/skeptical explanation of $G$ is 
$\Pi_2^P$-complete/$\Pi_3^P$-complete. 
\end{enumerate}
\item Given a ground negative observation $G$: 
\begin{enumerate}
\item Deciding if $G$ has a credulous/skeptical anti-explanation 
$\Sigma_2^P$-complete/$\Sigma_3^P$-complete. 
\item Deciding if an abducible $A\in {\cal A}$ is relevant to 
some credulous/skeptical anti-explanation 
(resp.\ some {\em minimal\/} credulous/skeptical anti-explanation) of $G$ 
is $\Sigma_2^P$-complete/$\Sigma_3^P$-complete 
(resp.\ $\Sigma_3^P$-complete/$\Sigma_4^P$-complete). 
\item Deciding if an abducible $A\in {\cal A}$ is necessary for every 
(minimal) credulous/skeptical anti-explanation of $G$ is 
$\Pi_2^P$-complete/$\Pi_3^P$-complete. 
\end{enumerate}
\end{enumerate}
In particular, when $P$ contains no disjunctive rules (i.e., $P$ is an ELP), 
the complexity of each problem decreases by one level in the polynomial 
hierarchy. 
\begin{proof}
1. For explaining positive observations, extended abduction 
includes normal abduction as a special case. 
Then, the hardness results of (a)--(c) hold by the corresponding 
decision problems of Proposition~\ref{complex-lem}. 
Since extended abduction is efficiently translated into 
normal abduction (Proposition~\ref{trans-xabd}), the membership results hold. 

2. Any credulous/skepcitcal anti-explanation of $G$ with respect to 
$\pf{P,{\cal A}}$ is 
equivalent to a credulous/skeptical explanation of $G'$ with respect to 
$\pf{P\,\cup\,\{\,G'\leftarrow not\,G\,\},{\cal A}}$ (Lemma~\ref{anti-lem}). 
Then, the results hold by the part 1 of this theorem. 
\end{proof}
\end{theorem}

The complexity results of extended abduction imply the complexity 
of view updates and theory updates. 
In what follows, we say that a view update or a theory update has a solution 
if there is an updated program which fulfills an update request. 

\begin{theorem}[complexity results for view updates]
Let $\pf{P,{\cal A}}$ be a propositional abductive program which 
represents a view update problem. Given a ground literal $G$: 
\begin{description}
\item[](a) Deciding if a view update has a solution in an EDP (resp.\ ELP) $P$ 
is $\Sigma_3^P$-complete (resp.\ $\Sigma_2^P$-complete) for inserting $G$, 
and $\Sigma_2^P$-complete (resp.\ NP-complete) for deleting $G$. 
\item[](b) Deciding if an abducible $A\in {\cal A}$ is relevant to 
a solution of a view update in an EDP (resp.\ ELP) $P$ 
is $\Sigma_4^P$-complete (resp.\ $\Sigma_3^P$-complete) for inserting $G$, 
and $\Sigma_3^P$-complete (resp.\ $\Sigma_2$-complete) for deleting $G$. 
\item[](c) Deciding if an abducible $A\in {\cal A}$ is necessary for 
every solution of a view update in an EDP (resp.\ ELP) $P$  
is $\Pi_3^P$-complete (resp.\ $\Pi_2^P$-complete) for inserting $G$, 
and $\Pi_2^P$-complete (resp.\ co-NP-complete) for deleting $G$. 
\end{description}
\begin{proof}
(a) Deciding the existence of a solution which accomplishes 
a view update for inserting (resp.\ deleting) $G$ is equivalent to 
the problem of deciding the existence of a skeptical explanation 
(resp.\ a credulous anti-explanation) of $G$ with respect to 
$\pf{P,{\cal A}}$ (Theorem~\ref{th-vu}). 
Hence, the result follows by Theorem~\ref{complexity-th}-1,2(a).
The results of (b) and (c) also follow from the corresponding 
decision problems of extended abduction of 
Theorem~\ref{complexity-th}-1,2(b),(c). 
\end{proof}
\end{theorem} 

\begin{theorem}[complexity results for theory updates]
Let $\pf{P,{\cal A}}$ be a propositional abductive program which 
represents a theory update problem. 
\begin{description}
\item[](a) Deciding if a theory update has a solution in an EDP (resp.\ ELP) 
$P$ is $\Sigma_2^P$-complete (resp.\ NP-complete). 
\item[](b) Deciding if an abducible $A\in {\cal A}$ is relevant to 
a solution of a theory update in an EDP (resp.\ ELP) 
$P$ is $\Sigma_3^P$-complete (resp.\ $\Sigma_2$-complete). 
\item[](c) Deciding if an abducible $A\in {\cal A}$ is necessary for 
every solution of a theory update in an EDP (resp.\ ELP) 
$P$ is $\Pi_2^P$-complete (resp.\ co-NP-complete). 
\end{description}
\begin{proof}
(a) Deciding the existence of a solution of a theory update is 
equivalent to the problem of deciding the existence of a credulous 
anti-explanation 
of $G=\bot$ with respect to $\pf{P,{\cal A}}$ (Theorem~\ref{th-tu}). 
Hence, the result holds by Theorem~\ref{complexity-th}-2(a).
The results of (b) and (c) also follow from the corresponding decision 
problems of extended abduction of Theorem~\ref{complexity-th}-2(b),(c). 
\end{proof}
\end{theorem}

\begin{cor}[complexity results for consistency restoration] 
Let $\pf{P,{\cal A}}$ be a propositional abductive program which 
represents an integrity maintenance (or inconsistency removal) problem. 
\begin{description}
\item[](a) Deciding if an integrity maintenance (or inconsistency removal) 
has a solution in an EDP (resp.\ ELP) is 
$\Sigma_2^P$-complete (resp.\ NP-complete). 
\item[](b) Deciding if an abducible $a\in {\cal A}$ is relevant to a solution 
of an integrity maintenance (or inconsistency removal) in an EDP (resp.\ ELP) 
is $\Sigma_3^P$-complete (resp.\ $\Sigma_2$-complete). 
\item[](c) Deciding if an abducible $a\in {\cal A}$ is necessary for every 
solution 
of an integrity maintenance (or inconsistency removal) in an EDP (resp.\ ELP) 
is $\Pi_2^P$-complete (resp.\ co-NP-complete). 
\end{description}
\begin{proof}
Since integrity maintenance or inconsistency removal is characterized 
as a special case of view deletion or theory update, 
the decision problems of these tasks have the same complexities as 
the corresponding problems of view deletion or theory update. 
\end{proof}
\end{cor}

\begin{table}
\caption{Complexity results for program updates}
\label{table-complex}
\begin{center}
\begin{minipage}{3.5in}
\begin{tabular}{lccc} \hline
Update EDP/ELP & existence & relevance & necessity \\ \hline\hline
view insertion & $\Sigma^P_3$/$\Sigma^P_2$ & $\Sigma^P_4$/$\Sigma^P_3$ & $\Pi^P_3$/$\Pi^P_2$ \\
view deletion  & $\Sigma^P_2$/NP & $\Sigma^P_3$/$\Sigma^P_2$ & $\Pi^P_2$/co-NP \\ \hline 
theory update & $\Sigma^P_2$/NP & $\Sigma^P_3$/$\Sigma^P_2$ & $\Pi^P_2$/co-NP \\ 
consistency restoration & $\Sigma^P_2$/NP & $\Sigma^P_3$/$\Sigma^P_2$ & $\Pi^P_2$/co-NP\\  \hline
\end{tabular}
\end{minipage}
\end{center}
\end{table}

The complexity results are summarized in Table~\ref{table-complex}. 
In the table, every entry represents completeness for the respective class. 
Also, consistency restoration means 
integrity maintenance or inconsistency removal. 

These complexity results show that decision problems for 
view insertion are generally harder than those of view deletion 
by one level of the polynomial hierarchy, while problems for 
theory updates and consistency restoration are as hard as 
those for view deletion. 

\section{Related work} \label{sec:7} 

There are a large number of studies which concern (normal) abduction and 
updates in logic programs and deductive databases.  In this section, 
we mainly discuss comparison with studies which handle {\em nonmonotonic\/} 
logic programs or deductive databases with negation. 

\subsection{Abduction} \label{sec:7.1}

There are a number of procedures for computing normal abduction. 
Studies \cite{EK89,KM90,Dec96,DS98} introduce top-down procedures for 
normal abduction. 
Top-down procedures efficiently compute abduction in a goal-driven manner, 
and the above procedures are correct for locally stratified NLPs. 
In unstratified programs, however, top-down (abductive) procedures are 
generally incorrect under the stable model semantics. 
By contrast, there exist correct top-down procedures in unstratified programs 
under different semantics.  For instance, 
Dung \shortcite{Dung91} shows that Eshghi and Kowalski's abductive 
procedure is correct with respect to the {\em preferred extensions}. 
Brogi $et\,al.$ \shortcite{BLM95} extend Kakas and Mancarella's 
procedure to extended logic programs, which works correctly under the 
{\em three-valued stable model semantics}. 
Alferes $et\,al.$ \shortcite{APS99} propose a tabled procedure for 
normal abduction under the {\em well-founded semantics}.  
These procedures compute positive explanations for positive observations in 
the context of normal abduction, while negative explanations 
or anti-explanations 
in extended abduction are not directly computed by these procedures. 
On the other hand, \cite{CDT91,FK97} provide bottom-up procedures 
which compute normal abduction through Clark's program completion. 
Using program completion, one can compute anti-explanations by treating the 
negative observation $p$ as $\neg p$ for the atom $p$ in the completion 
formula. However, these procedures are also restricted to programs where 
completion is well-defined.  

The approach taken in this paper is based on the computation 
of answer sets, which is executed in a bottom-up manner. 
This is the so-called {\em answer set programming\/} (ASP) 
which attracts much attention recently \cite{MT99,Nie99,Lif02}. 
Some researchers apply ASP to computing normal abduction. 
Inoue and Sakama \shortcite{IS96} introduce a procedure for normal abduction, 
which is based on the bottom-up fixpoint computation of extended disjunctive 
programs. 
Eiter $et\,al.$ \shortcite{EFL99} develop the system called ${\tt dlv}$ 
which has a front-end for abductive diagnoses in normal disjunctive 
programs.  These two studies use program transformations from 
abductive programs to disjunctive programs and find 
credulous (minimal) explanations of normal abduction using 
bottom-up computation of answer sets or stable models. 
This paper introduced a program transformation from abductive programs 
to update programs, but it is different from these studies in the following 
points.  First, update programs are prepared for extended abduction 
and can compute negative (anti-)explanations as well as positive ones. 
Second, we provide methods of computing both credulous and 
skeptical (minimal) explanations through update programs. 
Satoh and Iwayama \shortcite{SI91} provide a transformation from abductive 
programs to normal logic programs under the stable model semantics, and 
Toni and Kowalski \shortcite{TK95} provide another transformation 
under the {\em argumentation framework}.  
These transformations are applied to normal abduction and do not 
consider disjunctive programs. 

Update programs are simple and applied to a broader class of abductive 
programs, and extended abduction is realized by any procedure for computing 
answer sets of EDPs.%
\footnote{In implementation, some restrictions on programs 
such as function-free and range-restricted conditions would be necessary.} 
Moreover, 
since extended abduction includes normal abduction as a special case, 
update programs are also used for computing normal abduction in EDPs. 
To compute extended abduction, \cite{IS99} introduced 
a {\em transaction program\/} which is a set of production rules to 
compute (anti-)explanations by fixpoint construction. 
The procedure works correctly in acyclic covered NLPs. 
Inoue \shortcite{Ino00} introduces a simple program transformation 
from extended abduction to normal abduction in general EDPs. 
Using the transformation, extended abduction is executed via 
normal abduction. 

\subsection{View update} \label{sec:7.2} 

In deductive databases, update requests on view definitions are considered 
observations and extensional facts are identified with abducible hypotheses. 
Then, abduction is viewed as the process of identifying possible changes 
on extensional facts. 
Early studies which realize view updating through abduction are 
based on this idea \cite{KM90,Bry90,CSD95,Dec96}. 
However, existing approaches characterize the view update problem using 
normal abduction, which result in somewhat indirect formulations for 
representing fact removal or view deletion. 
For instance, Kakas and Mancarella \shortcite{KM90} realize the deletion 
of a fact $A$ by the introduction of 
a new atom $A^*$ which represents $not\,A$ 
together with the integrity constraints $\leftarrow A,A^*$ and $A\vee A^*$. 
Bry \shortcite{Bry90} specifies the deletion of a fact $A$ using the 
meta-predicate $new(\neg A)$. Console $et\,al.$ \shortcite{CSD95} realize 
the deletion of a fact $A$ by the insertion of $\neg A$ under 
program completion. 
Bry and Console $et\,al.$ handle normal logic programs, so that this 
conversion causes no problem.  However, deleting $A$ and inserting $\neg A$ 
have different effects 
when the background program contains negative facts explicitly as in 
extended logic programs. 

Procedurally, \cite{KM90,CSD95} separate the process of view updating 
into two-steps; 
computing abductive explanations in the intensional database and 
updating base facts in the extensional database. 
Such a separation is effective to reduce the cost of 
extensional database accesses, 
while it does not reflect the current state of the extensional 
database and may lead to redundant computation. 
For instance, consider the program: 
\begin{eqnarray*}
&& p\leftarrow a_1,\,b,\\
&& \cdots\\
&& p\leftarrow a_k,\,b,\\
&& a_1\leftarrow,\;\ldots,\; a_k\leftarrow.
\end{eqnarray*}
where $a_i\;(i=1,\ldots,k)$ and $b$ are extensional facts. 
Given the update request to insert $p$, 
Kakas and Mancarella \shortcite{KM90} and 
Console $et\,al.$ \shortcite{CSD95} compute $k$ minimal explanations 
$\{a_1,b\},\ldots,\{a_k,b\}$ in the intensional database, 
which are evaluated in the extensional database. 
Such computation is unnecessary and ineffective, 
since $\{b\}$ is the unique minimal explanation of $p$. 
Decker \shortcite{Dec96} introduces an improved version of the abductive 
procedure which avoids such redundant computation by including base facts 
in the input of refutation processes. 
However, \cite{Dec96} does not take base facts into account 
during the consistency derivations.  As a result, it often fails to obtain 
correct solutions \cite{MaT99}. 
Abductive procedures of \cite{KM90,Dec96} 
are top-down and the correctness is guaranteed for locally stratified NLPs. 
Similarly, view updating based on SLDNF-like top-down procedures such 
as \cite{Dec90,GuL90,GuL91,TO95} have restrictions on the program syntax. 
By contrast, our method is based on the computation of answer sets, 
which is executed in a bottom-up manner and is applicable to any EDP. 
Bry \shortcite{Bry90} and Console $et\,al.$ \shortcite{CSD95} also compute 
view updates in a bottom-up manner. 
The former specifies update procedures in a meta-program 
and the latter uses Clark's completion.  
They realize view updates in normal logic programs and 
do not handle disjunctions nor explicit negation in a program. 
For updating disjunctive programs, \cite{GHLM93,FGM96} provide algorithms 
for view updates in propositional NDPs.  
The former provides a top-down algorithm to compute view updates 
in stratified disjunctive programs, while the latter achieves view updates 
in NDPs by bottom-up computation.  Fernandez $et\,al.$ \shortcite{FGM96} 
first compute all possible models from the Herbrand base of extensional 
facts, then minimal models that satisfy updates are constructed from those 
models.  By contrast, we compute the answer sets of an update program 
and select the U-minimal ones, which is usually a much smaller set and 
is easier than \cite{FGM96}. 

In deductive databases, integrity maintenance is often coupled with 
view updating \cite{MaT99}. 
Concerning studies which handle integrity maintenance in 
nonmonotonic logic programs, 
\cite{TO95,Dec96} merge transactions of view updates 
and integrity maintenance in SLDNF-like top-down procedures. 
These procedures are sound 
and \cite{TO95} is also complete for computing view updates 
satisfying integrity constraints in locally stratified logic programs. 
Abductive procedures in \cite{KM90b,BLM95,TK95} also check integrity 
constraints in the process of computing candidate hypotheses. 
Compared with these studies, our approach in Section~\ref{sec:4.2} 
is based on the computation of answer sets and is applicable to 
non-stratified, disjunctive, and extended logic programs. 

\subsection{Theory update} \label{sec:7.3} 

Fagin $et\,al.$ \shortcite{FUV83} formalize theory updates for 
inserting/deleting a single sentence to/from a first-order theory. 
According to their definition, a theory $T$ accomplishes the insertion of 
the sentence $\sigma$ if $\sigma\in T$, while $T$ accomplishes the 
deletion of $\sigma$ if $\sigma\not\in Th(T)$ where $Th(\cdot)$ 
is the set of sentences proved by $T$. 
These definitions of insertion and deletion are not symmetric, i.e., 
derived sentences are taken into consideration in deletion, 
while they are not considered in insertion. 
In fact, if deletion is defined by $\sigma\not\in Th(T)$, 
it seems natural to define insertion also by $\sigma\in Th(T)$. 
In this paper, we achieve updates on derived facts by view updates, while 
explicit insertion/deletion of a sentence itself is distinguished as 
theory updates in Section~\ref{sec:5.1}.  Our Definition~\ref{df-ruleup} is 
symmetric for insertion and deletion of sentences. 
Fagin $et\,al.$'s update semantics is also characterized by extended 
abduction in \cite{IS95}, hence it is computable via update programs. 
In \cite{FKUV86} they extended the framework to updating a theory by 
several sentences. 
The definition of their {\em batch insertion\/} is close to 
the definition of our theory update of Definition~\ref{df-th-up}. 
A difference is that they handle first-order theories, while we consider 
nonmonotonic logic programs. Moreover, they provide no computational method 
to realize theory updates. 

Alferes $et\,al.$ \shortcite{AL00} introduce the framework of 
{\em dynamic logic programming\/} 
which realizes theory updates in nonmonotonic logic programs. 
They represent updates using meta-rules which specify changes between 
different states, and the result of update is reflected by the stable 
models of the updated program. 
Compared with our framework, \cite{AL00} computes stable models of an 
updated program but does not compute an updated program at the object level. 
Moreover, the effect of updates is also different from ours. 
In Example~\ref{program-ex}, updating $P_1$ with a series of updates 
$P_2$ and $P_4$ results in the program 
which has the answer set $\{\,\neg power\_failure,\, sleep\,\}$. 
Interestingly, however, starting from the same knowledge base 
and applying the same updates,%
\footnote{\cite{AL00} uses default negation $not$ instead of 
explicit negation $\neg$ in the head of rules.} 
\cite{AL00} revives the original program 
$P_1$ and concludes $\{\,tv\_on,\, watch\_tv\,\}$. 
Thus, after power is back up again, TV automatically works 
and a person watches TV in Alferes $et\,al.$'s approach, 
while this is not the case in our semantics. 
This difference comes from the fact that \cite{AL00} considers that 
every rule/fact in the initial program $P_1$ persistently holds unless 
it is forced to be false by updates.  
Besides, persistent sentences once rejected by an update revive when 
the update is later invalidated.  
However, such a persistent assumption works too strong in many situations.%
\footnote{They call it the ``principle of inertia'' but the assumption is 
stronger than the law of inertia in the usual sense.} 
For instance, consider the following scenario. 
``A person planned to go to a concert on this Friday's evening and reserved a 
seat. After a while, however, a meeting was scheduled with his client on 
that day, so he canceled the reservation. 
On Friday morning, there is a call from the client that she will be absent 
from the meeting because of illness.'' 
The situation is described as follows. 
The initial situation is 
\[ Q_1:\; seat\_reserved\leftarrow. \]
Updating $Q_1$ with 
\begin{eqnarray*}
Q_2:&& \neg seat\_reserved\leftarrow cancel\_reservation,\\
&& cancel\_reservation\leftarrow meeting\_scheduled,\\
&& meeting\_scheduled\leftarrow
\end{eqnarray*}
amounts to the new program $Q_2$. 
Next, updating $Q_2$ with 
\begin{eqnarray*}
Q_3:&& \neg meeting\_scheduled\leftarrow meeting\_canceled,\\
&& meeting\_canceled\leftarrow client\_absent,\\
&& client\_absent\leftarrow, 
\end{eqnarray*}
results in the program 
$(Q_2\cup Q_3)\setminus \{\,meeting\_scheduled\leftarrow\,\}$, 
which has the answer set 
$\{\,client\_absent,\,meeting\_canceled,\,\neg meeting\_scheduled\,\}$. 
On the other hand, according to \cite{AL00}, the fact in $Q_1$ 
revives after updating $Q_2$ with $Q_3$. 
As a result, it automatically recovers cancelled reservation 
after the client's call, which is unintuitive. 
In real life, once a state has changed by an update, the state is not 
always recovered again just by cancelling the effect of an update. 
A knowledge base generally contains persistent knowledge and 
temporary knowledge, and it is important to distinguish them. 
Back to the TV example, if $tv\_on$ holds by default whenever the 
power is supplied, it is represented as a default rule
\[ tv\_on\leftarrow not\,power\_failure\,. \] 
In this case, we have the same result as \cite{AL00} after updates. 

Alferes $et\,al.$ \shortcite{AP99} propose a language called 
{\em LUPS\/} for specifying changes to logic programs. 
It realizes a theory update by a series of update commands which are 
translated into a normal logic program written in a meta-language 
under the stable model semantics. 
Using LUPS, persistent/non-persistent rules are distinguished 
by the commands ${\tt always}$/${\tt assert}$, which are 
respectively cancelled by ${\tt cancel}$/${\tt retract}$. 
For instance, the situation in the above example is expressed as 
\begin{eqnarray*}
&& {\tt assert}\; seat\_reserved.\\
&& {\tt retract}\; seat\_reserved\; {\tt when}\; meeting\_scheduled.
\end{eqnarray*}
Then, reservation is cancelled when meeting is scheduled, and 
the reservation is never recovered just by cancelling the meeting. 
Compared with their approach, update programs considered in this paper 
specify changes at the object level.  Moreover, our update programs are 
used for computing not only theory updates but also view updates. 

Eiter $et\,al.$ \shortcite{EFS00} reformulate the approach of \cite{AL00} 
and introduce update programs which have the same effect as dynamic 
logic programs. 
They also introduce minimal and strict updates 
in an update sequence.  For instance, consider the program sequence: 
\begin{eqnarray*}
P_1:&& a\leftarrow,\\
P_2:&& \neg a\leftarrow not\,c,\\
P_3:&& c\leftarrow not\,d,\;\; d\leftarrow not\,c. 
\end{eqnarray*}
First, updating $P_1$ by $P_2$ has the single answer set $\{\neg a\}$ as 
the solution. Next, updating $P_2$ by $P_3$ has 
two answer sets $S_1=\{c\}$ and $S_2=\{\neg a,d\}$. 
Among these two, $S_1$ is consistent with $P_1\cup P_2$, 
while $S_2$ is inconsistent with $P_1$ then $P_1$ is rejected. 
In this case, $S_1$ is called {\em minimal\/} with respect to 
historical changes, and is preferred to $S_2$. 
A {\em strict\/} update further takes the temporal order of updates into 
consideration.  Our theory updates just consider the minimal change between 
the current knowledge base and the new one, and do not take the history of 
updates into consideration.  
In the above example, our theory updates produce 
the program $P_2\cup P_3$ and both $S_1$ and $S_2$ are the solutions. 
However, the selection of minimal updates with respect to historical changes 
is not always intuitive.  For instance, consider the scenario: 
First, a person planned to join a party as she had no schedule on that day 
($P_1$).  After a while, she got a job which must be done by that day. 
If she is not free due to the job, she cannot join the party ($P_2$).  
Now, the party is tomorrow.
But she does not know whether she can finish the job before the party ($P_3$).
The scenario is represented by the program sequence: 
\begin{eqnarray*}
P_1:&& join\_party\leftarrow,\\
P_2:&& \neg join\_party\leftarrow not\,free,\\
P_3:&& free\leftarrow not\,busy,\\
    && busy\leftarrow not\,free,
\end{eqnarray*}
which have the same structure as the preceding example. 
In this case, there seems no reason to prefer 
$\{\,join\_party,\,free\,\}$ to $\{\,\neg join\_party,\,busy\,\}$, 
since according to the latest information $P_3$ 
it is not known whether $free$ or $busy$. 
Eiter $et\,al.$'s approach is based on the {\em causal rejection principle\/} 
which states that an old rule $r$ is discarded by a more recent rule $r'$ 
only if $r$ contradicts $r'$. 
The causal rejection principle resolves contradiction 
between old and new programs, 
but does not resolve contradiction which arises in a program. 
For instance, updating the program 
$P_1=\{\,q\leftarrow,\;\; \neg q\leftarrow a\,\}$ 
with $P_2=\{\,a\leftarrow\,\}$ has no solution by the causal rejection 
principle, 
while we have solutions by removing one of the two rules in $P_1$. 

Buccafurri $et\,al.$ \shortcite{BFL99} introduce an 
{\em inheritance program\/} which consists of a set of EDPs ordered by 
a generality relation. 
It realizes default reasoning in inheritance hierarchies 
and is also applied to updating logic programs. 
According to \cite{EFS00}, inheritance programs are equivalent to 
update programs of Eiter $et\,al.$'s, hence the same arguments as the 
comparison with update programs are applied. 

Zhang and Foo \shortcite{ZF98} study theory updates between ELPs. 
When updating $P_1$ with $P_2$, 
they first update each answer set $S$ of $P_1$ with $P_2$. 
The result of this update, $S'$, is a set of ground literals which 
has minimal difference from $S$ and satisfies each rule in $P_2$. 
Next, a maximal subset $P'\subseteq P_1$ is extracted such that 
$S'$ is a subset of an answer set of $P'\cup P_2$. 
When there is a conflict between rules in $P'$ and $P_2$, 
a higher priority is put on rules in $P_2$ and 
those rules are selected in the resulting program. 
Our theory update is different from theirs in both the method and the result. 
First, their update consists of a series of transactions: 
computation of the answer sets of the original program, 
updates on these answer sets, extraction of rules from the original program, 
merging two programs, and conflict resolution based on preference. 
By contrast, we perform a theory update in a much simpler manner 
by translating a program into an update program and computing the U-minimal 
answer sets of the update program. 
Second, conflict resolution taken in their approach often has an effect 
which seems too strong. 
As pointed out by \cite{EFS00}, updating $P_1=\{\,p\leftarrow not\,q\,\}$ 
with $P_2=\{\,q\leftarrow not\,p\,\}$ results in $P_2$, even though 
$P_1\cup P_2$ is consistent.  In our framework, the result of update is 
$P_1\cup P_2$. 

Decker \shortcite{Dec97} provides an abductive procedure for computing both 
{\em user updates\/} and {\em schema updates\/} in normal logic programs. 
User updates corresponds to view updates, while 
schema updates consider updating a theory with a rule. 
The procedure is top-down and works correctly for locally stratified 
programs. 
Studies \cite{BB95,Bou96,LU96} characterize belief update/revision based on 
normal abduction in monotonic propositional theories.  These approaches are 
the so-called ``interpretation updates'' and compute updates in terms of 
individual models of a theory.  
This is in contrast to our theory updates which 
computes updates directly by a program. 

To resolve inconsistency in a nonmonotonic logic program, 
Pereira $et\,al$. \shortcite{PAA91} introduce a method of contradiction 
removal in extended logic programs. When conflicting conclusions are 
brought by a program, they prefer a conclusion that does not depend on any 
default assumption.  This method does not resolve inconsistency in a 
program of Example~\ref{nixon-diamond}, 
where contradiction is brought by no default assumption. 
\cite{DP95} uses abduction to resolve inconsistency in ELPs. 
When a program derives contradiction, it is resolved by changing the 
truth value of abducible literals from true to false or undefined 
under the well-founded semantics. 
Yuan and You \shortcite{YY98} formalize the same problem 
by a three-valued semantics and resolve inconsistency in ELPs 
using a suitable program transformation. In their approach, 
the revised programs contain newly introduced literals. 
These studies use three-valued semantics and have different handling of 
inconsistency in general. 
For instance, the rule $p\leftarrow not\,p$ makes a program inconsistent 
under the answer set semantics, while $p$ is interpreted undefined 
under the well-founded semantics. 
Syntactically, the above studies do not handle programs containing 
disjunctions. 
Witteveen and van der Hoek \shortcite{WH97} consider a 
{\em back-up semantics\/} 
when the intended semantics fails to provide a consistent meaning to a 
program. For instance, when a program is inconsistent under the stable model 
semantics, they consider the minimal model semantics as a back-up semantics. 
Then, the program is made consistent by introducing some sentences which 
are supported by the back-up semantics. 
In this approach two different semantics are considered on the same program 
and the result of revision depends on the choice of a back-up semantics. 
Moreover, it does not resolve contradiction in the type of program of 
Example~\ref{nixon-diamond}. 

Inoue \shortcite{Ino94} characterizes inconsistency resolution 
in an ELP $P$ by the abductive program $\pf{\emptyset,P}$.  
Then, he considers a maximal consistent subset of the hypotheses $P$, 
which is computed using a program transformation from the abductive program 
to an ELP. 
We characterized the same problem by the abductive program $\pf{P,P}$ in 
Section~\ref{sec:5.3}, but the result is the same as \cite{Ino94} for ELPs. 
The problem is also characterized by the abductive program 
$\pf{P,{\cal L}_P}$ in \cite{IS95}.  
This formulation, however, produces different results in general. 
For instance, given the inconsistent program 
$P=\{\,\neg p\leftarrow,\;\; \leftarrow not\,p\,\}$, 
$\pf{P,{\cal L}_P}$ has the minimal explanation $(\{p\},\{\neg p\})$ which 
produces the updated program $\{\,p\leftarrow,\;\; \leftarrow not\,p\,\}$.  
On the other hand, $\pf{P,P}$ has the minimal explanation 
$(\{\},\{\leftarrow not\,p\})$ and the result of update is 
$\{\,\neg p\leftarrow\,\}$. 
Thus, $\pf{P,{\cal L}_P}$ permits the introduction of new facts 
as well as the deletion of facts to resolve inconsistency. 
Generally, permitting introduction of sentences increases 
the number of possible solutions. 
Nevertheless, this type of inconsistency resolution is also realized by 
computing consistent U-minimal answer sets of 
the update program of $\pf{P,{\cal L}_P}$. 

\subsection{Belief revision} \label{sec:7.4}

Update is often distinguished from {\em (belief) revision\/} \cite{KM91}. 
That is, update targets the problem of changing one's belief up to date 
when the (external) world changes.  
By contrast, revision handles the problem of modifying one's belief 
when new information about the static world is obtained 
(while the external world does not change). 
In this paper we handled the problem of view updates and theory updates, 
both of which are caused by the change of the external world in general. 
A question is then whether the present approach is also applicable to 
revision.  
Our position on this point is as follows. 
It is true that the distinction between update and revision is useful in 
some contexts, however, we do not consider that such a distinction is always 
possible. 
For instance, recall the bird-fly example in Section~\ref{sec:1.2}: 
\begin{eqnarray*}
&& flies(x)\leftarrow bird(x), not\,ab(x),\\
&& ab(x)\leftarrow broken\mbox{-}wing(x),\\
&& bird(tweety)\leftarrow,\\
&& broken\mbox{-}wing(tweety)\leftarrow.
\end{eqnarray*}
When we observe that tweety flies, the program is updated by 
deleting the fact $broken\mbox{-}wing(tweety)$, for instance.
Is this belief change is update or revision?
On one hand, it is considered that the external world has changed --
tweety has healed; on the other hand, it is considered that the external
world never changes, but the reasoner has a wrong (initial) belief --
$broken\mbox{-}wing(tweety)$.

As this example indicates, the same problem is captured from different 
viewpoints.  Only by observing new evidence, one cannot judge in general 
whether it comes from the change of the (external) world or not. 
Moreover, some researchers argue that revision is viewed as update of mental 
states \cite{VS94}. 
In this sense, we do not strictly distinguish update and revision 
in this paper. 

Katsuno and Mendelzon \shortcite{KM91} distinguish update and revision 
in the context of propositional theories, and introduce postulates to 
distinguish them. 
We do not examine these postulates in our update framework, 
but those postulates are defined for monotonic propositional theories 
and, as argued in \cite{EFS00}, they are not applicable to 
nonmonotonic updates in general. 
Katsuno and Mendelzon also argue that inconsistency in a knowledge base 
is resolved by revision rather than update. 
However, we often have inconsistent information in daily life, 
and resolve inconsistency by acquiring more accurate information. 
This process is captured as an update of one's mental state. 
Inconsistency removal considered in this paper is an example of this 
type of updates. 

\section{Conclusion}
This paper introduced an abductive framework for computing various update 
problems in nonmonotonic logic programs. 
The first contribution of this paper is a computational method for 
extended abduction through update programs. 
Update programs are extended disjunctive programs which are obtained by 
a simple program transformation from abductive programs. 
Then, (minimal) credulous/skeptical 
(anti-)explanations of positive/negative observations are 
computed by the (U-minimal) answer sets of an update program. 
The second contribution of this paper is characterizations of 
view updates and theory updates in terms of extended abduction. 
Extended abduction is suitable for formalizing information changes in 
nonmonotonic theories, 
and different types of updates are computed by the U-minimal answer sets of 
update programs in a uniform manner.  
Using update programs, computation of updates is realized on top of the 
existing procedures for answer set programming with the additional 
mechanism of selecting U-minimal answer sets. 

It has been widely recognized that abduction plays an important role in 
updating data and knowledge bases.  
The advantage of the present paper lies in its capability of uniform 
treatment of different types of theory changes 
as well as in its syntactic generality of the language. 
Formalizing various update problems in a single framework clarifies the 
difference of each update, and implies the possibility of integrating them. 
For instance, integrity maintenance and inconsistency removal 
are captured as special cases of view updates and theory updates, 
respectively. 
Then, consistency restoration is 
done as a sub-task of the corresponding update procedure. 
Further, it is possible to execute view updates and theory updates 
in a combined manner. For instance, suppose a knowledge base $K$ which 
consists of the invariable part $K_1$ and the variable part $K_2$. 
Then, an update on $K_1$ is done by view updates and an update on $K_2$ is 
done by theory updates. 
View updates and theory updates have been respectively studied in the 
field of databases and AI, but their combinations are not exploited in the 
literature due to different formulations. 
Thanks to the uniform treatment of this paper, 
we could provide a theoretical basis for such mixed types of updates. 

There is a trade-off between syntactic generality of the framework and 
the efficiency of the computational mechanism.  
Our abductive/update framework is general in the sense that it is 
applicable to any extended disjunctive program, while its computation is 
inefficient as it requires computing every answer set of an update program.  
As discussed in Section~\ref{sec:7.1}, goal-driven abduction does not produce 
correct answers in unstratified programs under the answer set semantics. 
On the other hand, the framework of extended abduction is independent of 
a particular semantics, so that abductive updates considered in this paper 
could be formulated under different semantics which has a correct top-down 
procedure. 
 From the complexity viewpoints, general update problems have 
very high complexity and are intractable in general (unless $P=NP$).  
Further, the update program $UP$ uses unstratified negation in $abd(\cdot)$, 
so that 
it is not evaluated efficiently even when the objective program $P$ is a 
stratified (normal) program. 
(When a program $P$ is a disjunctive program, 
replacing $abd(a)$ with the disjunctive fact $a;\overline{a}$ 
does not introduce 
unstratified negation to $UP$ as presented in Section~\ref{sec:3.1}.) 
One solution to avoid using unstratified negation is provided by 
\cite{Ino00} which introduces a simple translation from 
extended abduction to normal abduction.  
(The idea of this translation is presented in Section~\ref{sec:6}.) 
This translation keeps minimal (anti-)explanations, 
while it preserves the stratified structure of programs. 
An alternative formalization of update problems based on this transformation 
is left for future study.

\section*{Acknowledgments}
The authors thank the anonymous referees for their valuable comments.

\end{document}